\documentclass[11pt]{article}
\usepackage[german,english]{babel}
\usepackage[cp850]{inputenc}
\usepackage[dvips]{graphicx}
\usepackage{amsmath}
\usepackage{amsfonts}
\usepackage[usenames, dvipsnames]{color}
\usepackage{xcolor}
\usepackage{epsfig}

\font\tenmath=msbm10 scaled 1200

\font\sevenmath=msbm7 scaled 1200
\font\fivemath=msbm5 scaled 1200 

\newfam\mathfam \textfont\mathfam=\tenmath
\scriptfont\mathfam=\sevenmath \scriptscriptfont\mathfam=\fivemath
\def\math{\fam\mathfam}
\def\R{{\math R}}

\def\E{{\math E}}

\def\P{{\math P}}

\newtheorem{Thm}{Theorem}[section]
\newtheorem{Lem}{Lemma}[section]
\newtheorem{Pro}{Proposition}[section]
\newtheorem{Cor}{Corollary}[section]

\newfam\mathfam \textfont\mathfam=\tenmath
\scriptfont\mathfam=\sevenmath \scriptscriptfont\mathfam=\fivemath
\def\math{\fam\mathfam}

\def \^#1{\if#1i{\accent"5E\i}\else{\accent"5E#1}\fi}

\def \e{\varepsilon}
\def \g{\gamma}

\def \cqfd{\quad_\Box}
\def \ms{\medskip}
\def \ss{\smallskip}
\def \bs{\bigskip}
\def \ni{\noindent}

\title{\bf Optimal split of orders across liquidity pools: a stochastic algorithm approach}
 
\author{{\sc Sophie Laruelle} \thanks{Laboratoire de Probabilit\'es et 
Mod\`eles al\'eatoires, UMR~7599, Universit\'e Paris 6, 4, pl. Jussieu, F-75252 Paris Cedex 5, France. E-mail: {\tt  sophie.laruelle@upmc.fr}}  \and {\sc Charles-Albert Lehalle} \thanks{Head of Quantitative Research, Cr\'edit Agricole Cheuvreux, CALYON group ; 9 quai Paul Doumer, 92920 Paris La D\'efense. E-mail: {\tt clehalle@cheuvreux.com}} \and {\sc Gilles Pag\`es} \thanks{Laboratoire de Probabilit\'es et 
Mod\`eles al\'eatoires, UMR~7599, Universit\'e Paris 6, case 188, 4, pl. Jussieu, F-75252 Paris Cedex 5, France. E-mail: {\tt  gilles.pages@upmc.fr}}
}

\date{}

\begin{document}

\maketitle 

\begin{center}
First Version : 6th October 2009 \\
This Version : \today
\end{center}

\begin{abstract}
Evolutions of the trading landscape lead to the capability to exchange the same financial instrument on different venues. Because of  liquidity issues, the trading firms split large orders across several trading destinations to optimize  their execution. To solve this problem we devised two stochastic recursive learning procedures which adjust the proportions of the order to be sent to the different venues, one based on an optimization principle, the other on some reinforcement ideas. Both procedures are  investigated from a theoretical point of view: we prove $a.s.$ convergence of  the optimization algorithm under some light  ergodic (or ``averaging") assumption on the input data process. No  Markov property is needed. When the inputs are i.i.d. we show that the convergence rate is ruled by a Central Limit Theorem. Finally, the  mutual  performances of both algorithms are compared on  simulated and real data with respect to  an ``oracle" strategy devised by an  "insider" who   \textit{a priori} knows  the executed quantities by every venues.
\end{abstract}


\bs \ni {\em 2001 AMS classification:} 62L20,
secondary: 91B32,
62P05

\section{Introduction}
\setcounter{equation}{0}
\setcounter{Assumption}{0}
\setcounter{Theorem}{0}
\setcounter{Proposition}{0}
\setcounter{Corollary}{0}
\setcounter{Lemma}{0}
\setcounter{Definition}{0}
\setcounter{Remark}{0}

The trading landscape have seen a large number of evolutions following two regulations: Reg NMS in the US and MiFID in Europe. One of their consequences is the capability to exchange the same financial instrument on different trading venues. 
New trading destinations appeared to complement the trading capability of primary markets as the NASDAQ and the NYSE in the US, or EURONEXT, the London Stock Exchange and Xetra in Europe. Such alternative venues are called ``Electronic Communication Network'' (ECN) in the US and Multilateral Trading Facilities (MTF) in Europe. Each trading venue differentiates from the others at any time because of the fees or rebate it demands to trade and the liquidity it offers.

As the concerns about consuming liquidity increased with the financial crisis, trading firms use Smart Order Routers (SOR) as a key element in the process of optimizing their execution of large orders. Such devices are dedicated to split orders across trading destinations as a complement to the temporal slicing coming from the well known balance between the need to trade rapidly (to minimize market risk) and trading slow (to avoid market impact).

If the temporal slicing has been studied since the end of the nineties~\cite{
OPTEXECAC00} with recent advances to adapt it to sophisticated investment strategies~\cite{
citeulike:5094012}, this kind of spatial slicing (across trading destinations) has been mainly studied by economists from the point of view of its global market efficiency~\cite{
FOU06} rather than from one investor's point of view.

The complexity of spreading an order between $N$ trading destinations comes from the fact that you never knows the quantity $D_i$ available on the $i^{th}$ trading venue to execute your order of size $V$ during a time interval $\delta t$ at your given price. If the fraction $r_i\,V$ of your order that you sent to the $i^{th}$ liquidity pool is higher than $D_i$: you will loose time and may loose opportunity to execute $r_i\,V - D_i$ in an another pool; on another hand if $r_i\,V$ is lower than $D_i$: you will loose money if this pool fees are cheap, and an opportunity to execute more quantity here.
The only way to optimize such a split on real time is to adjust on the fly the proportions $(r_i)_i$ according to the result of your previous executions.

This paper is an in depth analysis of the optimal split of orders. The illustrations and most of the vocabulary come from the ``Dark pool'' case, where the price $S$ is not chosen by the trader (it is the market ``mid point'' price) and the answer of the pool is immediate (i.e. $\delta t=0$). Dark pools are MTFs that do not publish pre-trade informations, so an efficient use of the results of the previous executions (namely the realizations of the $\min(D_i^t,r_i^t\,V^t)$ for any $i$ and all $t$ in the past) is crucial.
The results exposed here solve the problem of simultaneously splitting orders and using the information coming back from the pools to adjust the proportions to send for the next order, according to a criteria linked to the overall quantity executed ($i.e.$ a linear combination of the $\min(D_i,r_i\,V)$).

The resulting trading strategy (which optimality is proven here) can be considered as an extension of the one conjectured by Almgren in~\cite{almsor08}. It may also be related to the class of multi-armed bandit recursive learning procedures, recently brought back to light in several papers (see~\cite{LAPATA,TAR}, \cite{LAPA1,LAPA2}; which in turn belongs to the wide  family  of ``recursive stochastic algorithms'' also known 
as ``stochastic approximation"  and extensively investigated in the applied 
probability literature (see~\cite{KUYI},~\cite{BEMEPR},~\cite{DUF}, etc)). 

In fact, we introduce two learning algorithms one based on an optimization under constraints principle  and a second algorithm based on a reinforcement principle for which we establish the existence of an equilibrium. We  extensively investigate the first one, considering successively the classical --~although unrealistic~-- case where the inputs (requests,  answers) are  i.i.d. and a setting in which the input only share some averaging properties. In the i.i.d. setting we establish $a.s.$ convergence of the procedure and a Central Limit Theorem relying on classical results from Stochastic Approximation Theory. By {\em averaging setting} (also referred as  {\em ergodic setting}), we mean that the inputs of the procedure has $a.s.$ an averaging property with respect to a distribution $\nu$ at a given rate, say $n^{-\beta}$, $\beta>0$, for a wide enough class of Borel functions. Typically, in our problem, these inputs are the successive $N+1$-tuples $(V^n,D^n_i,i=1,\ldots,N)$, $n\ge1$. Typically, if  we denote this input sequence  inputs by $(Y_n)_{n\ge 1}$, we will assume that, for every $f\!\in {\cal V}_{\beta,p}$,
\[
\frac1n\sum_{k=1}^n f(Y_k) -\int_{\R_+^{N+1}}fd\nu =O(n^{-\beta}) \quad \P\mbox{-}
a.s \mbox{ and in } L^p(\P).
\]
Usually, ${\cal V}_{\beta,p}$ is supposed to contain at least bounded continuous function $g:\R_+^{N+1}\to \R$ and subsequently all bounded $\nu$-$a.s.$ continuous functions. This will be enough for our purpose in this paper (Stochastic approximation in this general framework is investigate in~\cite{LAR}). 
But the key point to be noted here is that {\em no Markov assumption  is needed on this input sequence} $(Y_n)_{n\ge 1}$. These assumptions are hopefully  light enough to be satisfied by real data since it  can be seen as a kind of ``light" ergodicity at a given rate. In a Markovian framework it could be related to the notion of  ``stability" in the literature, see~\cite{DUF}.

Thus, this setting includes stationary $\alpha$-mixing processes (satisfying an Ibragimov condition) like those investigated in~\cite{DOU} (in~\cite{DEDetal} weaker dependence assumptions are made in the chapter devoted to stochastic approximation but the perturbation is supposed to be additive and non causal which is not at all the case in our problem). As concerns the second procedure for which no Lyapunov function seems to be (easily) made available, we establish the existence  of an equilibrium and show the $ODE$  related to the algorithm is a competitive system in the terminology  of monotonous differential systems extensively studied by Hirsch et al. (see $e.g.$~\cite{ANHI2}). The behaviour of such  competitive systems is known to be the most challenging, even when the equilibrium point is unique (which is not the case here).

Both procedures are compared in the final section, using simulated and real data. Further numerical tests and applications are ongoing works in CA Cheuvreux.

\ms
The paper is organized as follows: in Section~\ref{Deux}, we make precise the modeling of splitting orders among several venues in the framework of {\em Dark pools}, first in static then in a dynamic way. This leads to an optimization problem under constraints. In Section~\ref{Trois}, we study the execution function of one dark pool and introduce the recursive stochasic algorithm resulting from the optimization problem. In Section~\ref{Quatre} we analyze in depth this algorithm ($a.s.$ convergence and weak rate) when the ``innovations"  (data related to the orders, the executed quantities and the market price) are assumed i.i.d. In Section~\ref{Cinq} we extend the $a.s.$ convergence result to a more realistic framework  where these innovations are supposed to share some appropriate averaging properties ($e.g.$ satisfied by $\alpha$-mixing processes satisfying Ibragimov's condition).  Section~\ref{Six} is devoted to  the second learning procedure, based this time on reinforcement principle, introduced in~\cite{BELE}. We  make a connexion with the theory of  (competitive) monotonous dynamical systems. Finally, in Section~\ref{Sept}, we present several simulations results on simulated and real data to evaluate the performances of both procedures with respect to an ``oracle" strategy of an ``insider" who could know {\em a priori} the executed quantities by every dark pool.

 
\bigskip
 \noindent {\sc Notations:} $\bullet$ For every $N\ge 1$, set ${\cal I}_{_N}:=\{1,2,\ldots,N\}$, ${\cal P}_{_N} :=\{r=(r_i)_{1\le i\le n}\in \R_+^N\,|\, \sum_{i=1}^N r_i=1\}$. Let  ${\bf 1}^\perp :=\{u\!\in  \R^N\,|\, \sum_{i\in {\cal I}_{_N}}u_i=0\}$. 

  \noindent   $\bullet$ $\delta_{ij}$ denotes the Kronecker symbol.
 
 \noindent $\bullet$ $\langle\,.|\,.\rangle$  denotes the canonical inner product  on $\R^d$ and $|\,.\,|$ the derived  Euclidean norm. 

 \noindent $\bullet$ $int(A)$ denotes the interior of a subset $A$ of $\R^d$.
 
 \noindent $\bullet$  $\delta_a$ denotes the Dirac mass at $a\!\in \R^d$.
 
\section{A simple model for  the execution of orders by dark pools}\label{Deux}
\setcounter{equation}{0}
\setcounter{Assumption}{0}
\setcounter{Theorem}{0}
\setcounter{Proposition}{0}
\setcounter{Corollary}{0}
\setcounter{Lemma}{0}
\setcounter{Definition}{0}
\setcounter{Remark}{0}

\subsection{Static modelling} \label{Model}
As mentioned in the introduction, we will focus in this paper on the splitting order problem in the case of (competing)  {\em dark pools}. The execution policy of a dark pool  differs from a primary market: thus a dark pool   proposes  bid/ask prices  with no guarantee of executed quantity at the occasion of an over the counter transaction. Usually its bid price is lower  than the bid price offered on the regular market (and the ask price is higher). Let us temporarily focus on a buying order sent to several dark pools.   One can model the impact  of the existence of $N$  dark pools ($N\ge 2$) on a given transaction  as follows: let $V>0$ be  the random volume to be executed and let $\theta_i\!\in (0,1)$ be the {\em discount factor} proposed by the dark pool $i\!\in\{1,\ldots,N\}$. We will make the assumption that this discount factor is deterministic or at least known prior to the execution.  Let $r_i$ denote the percentage of $V$ sent to the dark pool $i$ for execution and let $D_i\ge 0$ be the quantity of securities that can be delivered (or made available) by the dark pool $i$ at price $\theta_i S$ where $S$ denotes the bid price on the primary market (this is clearly an approximation since on the primary market, the order will be decomposed into slices executed at higher and higher prices following the order book). The rest of the order has to be executed on the primary market, at price $S$. Then the  cost $C$ of the executed order is given by 
\begin{eqnarray*}
C&= &S\sum_{i=1}^N \theta_i \min(r_iV,D_i) + S\left(V-\sum_{i=1}^N \min(r_iV,D_i)\right)\\
&=& S\left(V- \sum_{i=1}^N \rho_i \min(r_iV,D_i)\right)
\end{eqnarray*}
where $\rho_i =1-\theta_i>0$, $i=1,\ldots,N$. 
At this stage,  one may wish to minimize the mean execution cost $C$, {\em given the price $S$}. This amounts to solving  the following (conditional)  maximization problem
\begin{equation}\label{Max:1}
\max\left\{\sum_{i=1}^N \rho_i\,\E \left( \min(r_iV,D_i)\,|\, S \right),\; r\!\in {\cal P}_{_N}\right\}.
\end{equation}
However, none of the agents being insiders, they do not know the price $S$ when the agent decides to buy the security and when the dark pools answer to their request. This means that one may assume that $(V,D_1,\ldots,D_n)$ and $S$ are independent so that  the maximization problem finally reads
\begin{equation}\label{Max:2}
\max\left\{\sum_{i=1}^N \rho_i\E\left( \min(r_iV,D_i)\right),\; r\!\in {\cal P}_{_N}\right\}
\end{equation} 

\noindent where we assume that   all the random variables  $\min(V,D_1)$, \dots, $\min(V,D_{_N})$ are integrable (otherwise the problem is meaningless). 

An alternative choice could  be to include the price $S$ of the security into the optimization which leads to the mean maximization problem
\begin{equation}\label{Max:3}
\max\left\{\sum_{i=1}^N \rho_i\,\E \left( S\min(r_iV,D_i)\right),\; r\!\in {\cal P}_{_N}\right\}
\end{equation}
(with the appropriate integrability assumption to make the problem consistent).  It is then convenient to {\em include the price $S$ into both random variables $V$ and $D_i$}  by considering $\widetilde V:=V\,S$ and $\widetilde D_i:=D_iS$ instead of $V$ and $D_i$ which leads again to the maximization problem~(\ref{Max:2}).

%

\smallskip
If one considers   symmetrically a selling order  to be executed, the dark pool is supposed to propose a higher  ask price $\theta_iS$, $\theta_i>1$, than the order book. The seller  aims at maximizing the execution global (mean) price of the transaction. This yields to the same formal optimization problem, this time with $\rho_i=\theta_i-1$, $i=1,\ldots,N$.

\medskip  All these considerations lead us to focus on the abstract optimal allocation problem~(\ref{Max:2}) which explains why the price variable $S$ will no longer appear explicitly in what follows.

\subsection{The  dynamical aspect}
In practice, there is no {\em a priori} assumption --~or information available~-- on the joint distribution of $(V,D_1,\ldots,D_{_N})$  under $\P$.
So the only  reasonable way to provide a procedure to solve  this allocation problem is to devise an {\em on-line learning  algorithm based on historical data}, namely the  results of former  transactions with the dark pools on this security executed  in the past. This underlines  that our  agent dealing with the dark pools is a financial institution like a bank,  a broker or possibly a large investor which often --~that means at least daily~-- faces some large scale execution problems on the same securities.

This means that we will have to make some assumptions on the dynamics  of these transactions $i.e.$ on the data input  sequence $(V^n,D_1^n,\ldots,D_{_N}^n)_{n\ge 1}$  supposed to be defined on the same probability space $(\Omega,{\cal A},\P)$.  

\medskip
Our basic assumption on the sequence $(D^n_i,V^n,i=1,\leq,N)_{n\ge 1}$ is of statistical ~--~or ergodic~--~nature: we ask this sequence to be $\nu$-averaging ($a.s.$ and in $L^p(\P)$), at least on bounded continuous functions, where $\nu$ is a distribution on $(\R_+^{N+1},{\cal B}or(\R_+^{N+1}))$.   This leads to the following formal assumption:  

\bigskip
\noindent $(ERG)_{\nu}\;\equiv\;$ $\displaystyle \left\{\begin{array}{cl}
(i) 
& \mbox{the sequence  $(V^n,D^n_i, i=1,\ldots,N)_{n\ge 1}$  is averaging $i.e.$}\\ 
&\displaystyle \P\mbox{-}a.s. \quad\frac 1n \sum_{k=1}^n \delta_{(V^k,D_1^k,\ldots,D^k_N)}\stackrel{(\R_+^{N+1})}{\Longrightarrow} \nu, \\
(ii) & \sup_n \E (V^n)^2<+\infty. 
\end{array}\right.$

\bigskip
\noindent where 
$\displaystyle \stackrel{(\R_+^{N+1})}{\Longrightarrow} $ denotes the weak convergence of probability measures on $\R_+^{N+1}$. For convenience, we will denote $(V,D_1,\ldots,D_{_N})$ the canonical random vector on $\R_+^{N+1}$ so that we can write $\nu={\cal L}(V,D_1,\ldots,D_{_N})$.

Assumption $(ii)$ on the marginal distribution of the sequence $(V^n)_{n\ge1}$ is mainly technical. In fact standard arguments from weak convergence theory show that  combining $(i)$ and $(ii)$ implies
\[
\frac 1n \sum_{k=1}^n V^k \longrightarrow \E\, V   \quad\mbox{ as }\quad n\to \infty
\]
($\sup_n \E (V^n)^{1+\varepsilon} <+\infty$ would be enough). An important subcase is the the  $(IID)$ setting 

\medskip
\noindent $(IID) \;\equiv\;$  $\displaystyle \left\{\begin{array}{cl}(i)& \mbox{the sequence  $(V^n,D_1^n,\ldots,D_{_N}^n)_{n\ge 1}$ is i.i.d. with distribution $\nu={\cal L}(V,D_1,\ldots,D_{_N})$},\\
(ii) & \mbox{$V\!\in L^{2}(\P)$}.
\end{array}\right.$

\bigskip
This more restrictive assumption is undoubtedly less realistic from a  modeling point of view but it remains acceptable as a first approximation. It is the most common framework  to apply the standard Stochastic Approximation machinery  ($a.s.$ convergence,  asymptotically normal fluctuations, etc). So, its interest may be considered at least as pedagogical. The $(ERG)$ setting is slightly more demanding in terms of assumptions and needs  more specific methods of proof. It will be investigated as a second step, using some recent results established in~\cite{LARPAG} which are well suited to the specificities of our  problem (in particular we will not need to assume the existence of a  solution to the Poisson equation related to the procedure like in the reference book~\cite{BEMEPR}).




\section{Optimal allocation: a stochastic Lagrangian algorithm}\label{Trois}
\setcounter{equation}{0}
\setcounter{Assumption}{0}
\setcounter{Theorem}{0}
\setcounter{Proposition}{0}
\setcounter{Corollary}{0}
\setcounter{Lemma}{0}
\setcounter{Definition}{0}
\setcounter{Remark}{0}

 \subsection{The mean execution function of a dark pool}
In view of the modeling section, we need to briefly describe the precise behaviour of the  mean execution function $\varphi:[0,1]\to \R_+$ of a single dark pool. 

Let $(V,D)$ be an $\R_+^2$-valued random vector defined on a probability space $(\Omega,{\cal A},\P)$ representing the {\em global volume}  to be executed and the {\em  deliverable quantity} (by the dark pool) respectively. Throughout this paper we will assume the following consistency assumption 
\begin{equation}\label{varphiL1}
V>0\quad\P\mbox{-}a.s. \quad \mbox{and}\quad \P(D>0)>0.
\end{equation} 

The $a.s.$ positivity of $V$ means that we only consider  true orders. The fact that $D$ is not identically $0$ means that the dark pool does exist in practice. The ``rebate" coefficient $\rho $ is specific to the dark pool.

To define in a consistent way the mean execution function of a dark pool we only need to assume that $V\in L^1(\P)$ (although more stringent integrability assumptions are made throughout the paper).

Here the  {\em mean execution function} $\varphi:[0,1]\to \R_+$ of the dark pool is defined by
\begin{equation}\label{varphi}
\forall\, r\!\in [0,1],\qquad \varphi(r) = \rho\,\E (\min(rV,D))
\end{equation}
where $\rho>0$. The function $\varphi$ is finite, non-identically $0$. It is   clearly a concave non-decreasing bounded function. Furthermore, one easily checks that  its right and left derivatives are given at every $r\!\in [0,1]$ by
\begin{equation}\label{Derivphi}
\varphi'_l(r)=  \rho\,\E\left(\mbox{\bf 1}_{\{r V\le  D \}}V\right)\qquad \mbox{ and }\qquad \varphi'_r(r)= \rho\, \E\left(\mbox{\bf 1}_{\{rV< D\}}V\right).
\end{equation}
In particular, 
$$
\varphi'(0)= \rho \,\E ( V\mbox{\bf 1}_{\{D>0\}})>0
$$
and if 

\begin{equation}\label{noatom}
\hbox{{\em the (right continuous) distribution function of $\frac {D}{V}$ is continuous on $\R_+$}},
\end{equation}
 
 \smallskip
 \noindent then 
 \[
 \hbox{{\em$\varphi$ is everywhere differentiable on the unit interval $[0,1]$ with $\varphi'=\varphi'_l$ on $(0,1]$}}.
 \]

Assumption~(\ref{noatom}) means  that the distribution of $\frac D V$ has no atom except possibly at $0$. It can be interpreted as the fact that a dark pool has no ``quantized" answer to an order.

More general models of execution functions in which the rebate $\rho$ and the deliverable  quantity $D$ may depend upon the quantity to be executed $rV$  are briefly discussed further on.


 \subsection{Design of the stochastic Lagrangian algorithm}
 Let $V$ be the quantity to be executed by $N$ dark pools. For every dark pool $i\!\in {\cal I}_{_N}$ the  available quantity $D_i$ is defined on the same probability space   $(\Omega,{\cal A}, \P)$ as $V$. We assume that all couples $(V,D_i)$ satisfy the consistency  assumption~(\ref{varphiL1}). 
 
 To each dark pool $i\!\in {\cal I}_{_N}$ is attached a (bounded concave) mean execution function $\varphi_i$ of type~(\ref{varphi}),  introduced in  Section~\ref{Model}, or~(\ref{varphi2}),  (\ref{varphi3}) studied in  Section~\ref{Huit}.

\smallskip
Then for every $r=(r_1,\ldots,r_{_N})\!\in {\cal P}_{_N}$, 
 \begin{equation}\label{MaxPhi}
 \Phi(r_1,\ldots,r_{_N}):=\sum_{i=1}^N \varphi_i(r_i).
 \end{equation}

 In order to design the algorithm  we will need to extend the mean execution function $\varphi$ (whatever its form is) as a concave function on the whole real line by setting
\begin{equation}\label{Extension}
\varphi(r)=\left(r- \frac{r^2}{2}\right)\varphi'(0)\quad \mbox{if } r< 0 \quad \mbox{ and } \quad \varphi(r) = \varphi(1)+\varphi'(1)\log r \quad \mbox{ if } r>1.
\end{equation}

\smallskip
 Based on the extension of the functions $\varphi_i$ defined by~(\ref{Extension}), we can formally  extend $\Phi$ on the whole affine hyperplane spanned by ${\cal P}_{_N}$ $i.e.$
 $$
 {\cal H}_{_N}:=\{r\!\in \R^N\,|\, \sum_ir_i=1\}.
 $$

As announced, we aim at solving the following maximization  problem 
 \[
 \max_{r\in {\cal P}_{_N}} \Phi(r)  
 \]
but we will also have to deal for algorithmic purpose  with the same maximization problem when $r$ runs over ${\cal H}_{_N}$.

  Before stating a rigorous result, let us a have a look at a Lagrangian approach that only takes into account  the  affine constraint that is $\displaystyle \max_{r} \Phi(r)-\lambda\sum_ir_i$. Straightforward formal computations suggest that 
  
  \bigskip
  \centerline{ $r^*\!\in {\rm argmax}_{{\cal P}_{_N}}  \Phi$  iff $\varphi'_i(r^*_i)$ is constant when $i$ runs over ${\cal I}_{_N}$}
  
  \bigskip
  \noindent or equivalently if
 
 \begin{equation}\label{CaracLag}
 \forall\, i\!\in {\cal I}_{_N},\qquad \varphi'_i(r^*_i) =\frac 1N \sum_{j=1}^N \varphi'_j(r^*_j).
 \end{equation}
 
In fact this statement is not correct in full generality because the Lagrangian method does not provide a necessary and sufficient condition for a point to be a maximum of a (concave) function; thus, it does not take into account the case where $\Phi$ reaches its maximum on the boundary  $\partial {\cal P}_{_N}$ where the above condition on the derivatives may fail. So, an additional assumption is necessary to make it true as established in the proposition below.

\begin{Pro}\label{MaxiPhi} Assume that $(V,D_i)$ satisfies the consistency assumptions~(\ref{varphiL1}) and~(\ref{noatom}) for every $i\!\in {\cal I}_{_N}$.

\medskip
\noindent $(a)$ Assume  that the functions $\varphi_i$ defined by~(\ref{varphi}) satisfy the following assumption
 \[
 ({\cal C})\quad \equiv \quad \min_{i\in {\cal I}_{_N}} \varphi'_i(0) \ge \max_{i\in {\cal I}_{_N}} \varphi'_i\left(\frac{1}{N-1}\right).
 \]
 Then  ${\rm argmax}_{{\cal P}_{_N}} \!\Phi$ is a compact convex set and 
\[
 {\rm argmax}_{{\cal P}_{_N}} \!\Phi = \{ r\!\in {\cal P}_{_N}, \,|\, \varphi'_i(r_i)= \varphi'_1(r_1),\; i=1,\ldots,N\}.
 \]
 Furthermore $ {\rm argmax}_{{\cal H}_{_N}} \!\Phi= {\rm argmax}_{{\cal P}_{_N}} \!\Phi$.

\medskip
\noindent $(b)$  If the functions $\varphi_i$ satisfy the slightly more stringent assumption, 
\[
 ({\cal C}_{<})\quad \equiv \quad\min_{i\in {\cal I}_{_N}} \varphi'_i(0) > \max_{i\in {\cal I}_{_N}} \varphi'_i\left(\frac{1}{N-1}\right).
 \]
 then 
 $$
 {\rm argmax}_{{\cal H}_{_N}} \!\Phi={\rm argmax}_{{\cal P}_{_N}} \!\Phi\subset int({\cal P}_{_N}).
 $$
 \end{Pro}

\noindent {\bf Remarks.} $\bullet$ If $N=2$, one checks that Assumption~(${\cal C}$) is also necessary to derive the conclusion of item~$(a)$.

\noindent $\bullet$ As a by-product of the proof below we have the following more precise result on the optimal allocation $r^*$: if $r^*\!\in {\rm argmax}_{{\cal P}_{_N}} $ and ${\cal I}_0(r^*):=\{i\!\in {\cal I}_{_N}\,|\, r^*_i=0\}$, then
\[
\max_{i\in {\cal I}_0(r^*)}\varphi'_i(0)
\le \min_{i\in {\cal I}_0(r^*)^c}\varphi'_i(0).
\]

\bigskip
\noindent {\sc Interpretation and comments:} $\bullet$ In the case of a ``regular"  mean execution function, Assumption~(${\cal C}$) is a kind of {\em homogeneity assumption on the rebates} made by the involved dark pools. If we assume that $\P(D_i=0)=0$ for every $i\!\in {\cal I}_{_N}$ (all dark pools  buy or sell at least one security with the announced rebate), then $({\cal C})$ reads
\[
\min_{i\in {\cal I}_{_N}} \rho_i \ge \max_{i\in {\cal I}_{_N}}\left( \rho_i\frac{\E\, V\mbox{\bf 1}_{\{\frac{V}{N-1}\le D_i\}}}{\E\, V}\right)
\]
since $\varphi'_i(0)= \rho_i \,\E \,V$. In particular, 
\[
\hbox{\em 
Assumption~(${\cal C}$)  is always satisfied when all the $\rho_i$'s are equal}
\]  
(all dark pools propose the same rebates).

\medskip 
\noindent $\bullet$ Assumption~(${\cal C}$) is in fact our main assumption in terms of modeling. It may look somewhat difficult to satisfy when the rebates are not equal. But the crucial fact in order  to preserve the generality of what follows is that it contains {\em no assumption about the dependence between the volume $V$ and the ``answers" $D_i$ from the dark pools}.

\bigskip
 \noindent {\bf Proof.}  $(a)$ The function $\Phi$ is continuous on a compact set hence ${\rm argmax}_{{\cal P}_{_N}} \!\Phi$ is not empty. Let $r\!\in{\rm argmax} _{{\cal P}_{_N}}\Phi$ and ${\cal I}_0(r):=\{i\!\in {\cal I}_{_N}\,|\, r_i=0\}$. Clearly ${\cal I}_0(r)\neq {\cal I}_{_N}$ so that  ${\rm card} \  {\cal I}_0(r)\le N-1$. Let $u\!\in {\bf 1}^\perp$ such that $u_i> 0$, $i\!\in {\cal I}_0(r)$. Then $t\mapsto \Phi(r+tu)$ defined on the right neighbourhood of $0$ reaches its maximum at $0$ so that its derivative at $0$ is non-positive. Specifying the vector $u$ yields
 
 \[
 \forall\, i\!\in {\cal I}_0(r),\;\forall\, j \!\in {\cal I}_0(r)^c,\quad \varphi_i'(0)\le\varphi'_j(r_j).  
 \]
 
Now if $u\!\in {\bf 1}^\perp$ with  $u_i= 0$, $i\!\in {\cal I}_0(r)$, then the $t\mapsto \Phi(r+tu)$ is defined on a neighbourhood of $0$ and reaches its maximum at $t=0$ so that its derivative is $0$ at $0$; specifying the vector $u$ yields

 \[
 \forall\, i,\, j\!\in {\cal I}_0(r)^c,\quad \varphi_i'(r_i)=\varphi'_j(r_j).  
 \]
 
Now, there exists at least one index   $i_1\!\in {\cal I}_0(r)^c$ such that $r_{i_1}\ge \frac{1}{|I_0(r)^c|}\ge \frac{1}{N-1}$. Hence $\varphi_{i_1}'(r_{i_1})\le \varphi_{i_1}'(\frac{1}{N-1})$ which implies in turn that for every $i_0\!\in {\cal I}_0(r)$, $\varphi'_{i_0}(0)\le\varphi_{i_1}'(r_{i_1}) \le    \varphi'_{i_1}(\frac{1}{N-1})$. Finally Assumption~$({\cal C})$ implies that these inequalities hold as equalities so that

\[
\forall\, i\!\in {\cal I}_{_N},\quad \varphi'_i(r_i)= \varphi'_1(r_1).
\]

Conversely, let $r\!\in {\cal P}_{_N}$ satisfying the above equalities. Then, for every $r'\!\in {\cal P}_{_N}$, the function $t\mapsto  \Phi(tr'+(1-t)r)$ is concave on $[0,1]$ with a right derivative equal to $0$ at $t=0$. So it is maximum at $t=0$ $i.e.$ $\Phi(r)\ge \Phi(r')$.

Now we pass to the the maximization over ${\cal H}_{_N}$. Since it is an affine space and   $\Phi$ is concave, it is clear, $e.g.$ by considering $\Phi$ as a function of $(r_1,\ldots,r_{N-1})$, that 
\[
 {\rm argmax}_{{\cal H}_{_N}} \!\Phi= \{ r\!\in {\cal H}_{_N}, \,|\, \varphi'_i(r_i)= \varphi'_1(r_1),\; i=1,\ldots,N\}
\]
(which is non-empty since it contains at least $ {\rm argmax}_{{\cal P}_{_N}} $).  Now let $r\!\in {\cal H}_{_N}\!\setminus {\cal P}_{_N}$. 
Assume there exists $i_0\!\in {\cal I}_{_N}$ such that $r_{i_0}<0$. Then  there always  exists an index $i_1\!\in {\cal I}_{_N}$ such that $r_{i_1}\ge \frac{1-r_{i_0}}{N-1}>\frac{1}{N-1}$. Consequently
\[ 
\varphi'_{i_0}(r_{i_0})=(1-r_{i_0})\varphi'_{i_0}(0)>\varphi'_{i_0}(0)\ge \min_i\varphi'_i(0)\ge \max_i \varphi'_i\left(\frac{1}{N-1}\right)\ge \varphi'_{i_1}\left(\frac{1}{N-1}\right)\ge \varphi'_{i_1}(r_{i_1})
  \]
which contradicts the equality of these two derivatives. Consequently all $r_i$'s are non-negative so that $r\!\in {\cal P}_{_N}$.

\medskip
\noindent $(b)$  If  ${\cal C}_{<}$ holds, the above proof shows that ${\cal I}_0(r)=\emptyset$ so that ${\rm argmax}_{{\cal P}_{_N}}\Phi_{_N}\subset int({\cal P}_{_N})$. $\cqfd$

 \subsection{Design of the stochastic algorithm}
 Now we are in position to devise the stochastic algorithm for the optimal allocation among the dark pools, taking advantage of the characterization of ${\rm argmax}_{{\cal P}_{_N}}\!\Phi$. In fact we will simply use the obvious remark that $N$  numbers  $a_1$,\dots, $a_{_N}$ are equal if and only if they are all equal to their arithmetic mean $\frac{a_1+\cdots+a_{_N}}{N}$. 
 
We consider the mean execution function as defined by~(\ref{varphi}). We assume from now on  that the continuity assumption~(\ref{noatom})  holds so that the representation~(\ref{Derivphi})  of its derivative can be taken as its right or its left derivative on $(0,1]$ (and its right derivative only at $0$).

 
 Using this representation~(\ref{Derivphi}) for all  the derivatives  $\varphi'_i$ yields that, if Assumption~$({\cal C})$ is satisfied, then $ {\rm argmax}_{{\cal H}_{_N}} \!\Phi= {\rm argmax}_{{\cal P}_{_N}} \!\Phi$ and 
   \[
r^*\!\in {\rm argmax}_{{\cal P}_{_N}} \!\Phi \Longleftrightarrow \forall\, i\!\in \{1,\ldots,N\},\; \E\left( V\left(\rho_i\mbox{\bf 1}_{\{r^*_i V\le   D_i \}}-\frac 1N \sum_{j=1}^N  \rho_j\mbox{\bf 1}_{\{r^*_j V\le   D_j \}}\right)\right)=0.
\]
However, the set ${\cal P}_{_N}$ is  not stable for the ``naive"  zero search algorithm  naturally derived from the above    characterization, we are led to devise the procedure on the hyperplane ${\cal H}_{_N}$.

Consequently, this leads to devise the following zero search procedure
\begin{equation} \label{LAlgo0}
r^{n} =r^{n-1}+\gamma_{n}H(r^{n-1}, V^{n},D_1^{n},\ldots,D^{n}_{_N}), \; n\ge 1, \quad r^0\!\in {\cal P}_{_N},
\end{equation}

\noindent where,  for every $ i\!\in {\cal I}_{_N}$, every 
$r\!\in{\cal H}_{_N}$, every $V>0$ and every  $D_1,\ldots,D_{_N}\ge 0$,  
\begin{eqnarray}\label{LAlgo}
  H_i (r,V, D_1,\ldots,D_{_N})&=&  V \Big (\rho_i\mbox{\bf 1}_{\{r_i V\le   D_i \}\cap \{r_i\in [0,1]\}}   - \frac 1N \sum_{j=1}^N  \rho_j \mbox{\bf 1}_{\{r_j V\le  D_j \} \cap \{r_j\in [0,1]\}}  \\
 \nonumber  && +R_i(r,V,D_1,\ldots,D_{_N} ) \Big)  
\end{eqnarray}
and the ``innovation"    $(V^n,D_1^n,\ldots,D_{_N}^n)_{n\ge 1}$ is a sequence of random vectors with non negative components such that, for every $n\ge 1$, $(V^n,D^n_i,i=1,\leq,N)\stackrel{d}{=}(V,D_i,i=1,\leq,N)$ and the remainder terms $R_i$ have a mean-reverting effect to pull back the  algorithm into ${\cal P}_{_N}$. They are designed from the extension~(\ref{Extension})  of the derivative functions $\varphi'_i$  outside the unit interval $[0,1]$; to be precise, for every $i\!\in {\cal I}_{_N}$,  
\begin{eqnarray*}
 R_i(r,V,D_1,\ldots,D_{_N})&=&\rho_i\left((1-r_i) \mbox{\bf 1}_{\{D_i>0\}\cap \{r_i<0\}}+\frac{1}{r_i}\mbox{\bf 1}_{\{V\le D_i\}\cap\{r_i>1\}}\right)\\
 && -\frac 1N \sum_{j=1}^N \rho_j\left((1-r_j) \mbox{\bf 1}_{\{D_j>0\}\cap \{r_j<0\}}+\frac{1}{r_j}\mbox{\bf 1}_{\{V\le D_j\}\cap\{r_j>1\}}\right).
 \end{eqnarray*}
 
  \subsection{Interpretation and implementability of the procedure}

\ni $\rhd$ {\sc Implementability.}  The vector $(r^n_i)_{1\le i\le N}$ in~(\ref{LAlgo0}) represents the dispatching of the orders among the $N$ dark pools  to be sent at time $n+1$ by the investor. It is computed at time $n$. On the other hand $V^n$ represents the volume to be executed (or its monetary value if one keeps in mind that we ``plugged" the price into the volume) and the $D^n_i$ the ``answer" of dark pool $i$, still at time $n$. 
 
 The point is that the investor does have no access to the quantities $D^{n}_i$. However, he/she knows what he/she receives from dark pool $i$, $i.e.$ $\min (D^n_i, r^{n-1}_iV^n)$. As a consequence, the investor has access to the event
 \[
 \{\min (D^n_i, r^{n-1}_iV^n)= r^{n-1}_iV^n\} = \{r^{n-1}_iV^n\le D^n_i\}
 \]
 
\ni which in turn makes possible the updating of the procedure although he/she has no access to  the true value of $D^n_i$.

So, except for edge effects outside the simplex ${\cal P}_{_N}$, the procedure as set can be implemented on real data.  

\ms
\ni $\rhd$ {\sc Interpretation.} As long as $r$ is a true allocation vector, $i.e.$  lies in the simplex ${\cal P}_{_N}$, the interpretation of the procedure is the following: assume first that all the factors $\rho_i$ are equal (to $1$). Then the dark pools which fully executed the sent orders ($r_iV\le D_i$)  are rewarded proportionally  to the numbers of dark pools which {\em did not fully executed} the request they received. Symmetrically, the dark pools which could not execute the whole request are penalized proportionally to the number of dark pools which satisfied the request.
  
  Thus, if only one dark pool, say dark pool $1$, fully executes the request at time $n$, its pourcentage will be increased for the request at time $n+1$ by $\gamma_{n}(1-\frac 1N)V^n$ $i.e.$ it will asked to execute $r^{n}_1 = r^{n-1}_1+ \gamma_{n}(1-\frac 1N)V^n$ \% of the total order $V^{n+1}$. The other  $N-1$ dark pools will be penalized symmetrically: the pourcentage $r^n_i$ of the total request $V^{n+1}$ each of them  will receive at time $n+1$ will be reduced by $\gamma_{n}\frac 1NV^n$.
  
 \ss
  If   $k$ dark pools totally execute their request at time $n$ and the $N-k$ other fail, the pourcentages of $V^{n+1}$ that  the ``successful"  dark pools will receive for execution at time $n+1$  will be increased by  $\gamma_{n}(1-\frac kN)V^n$, each of the $N-k$ ``failing dark pools" being reduced by $\gamma_{n}\frac kNV^n$ . 
  
 \ss
  If no dark pool was able to satisfy their received request at time $n$, none will be penalized and if all dark pools fully execute the received orders, none will be rewarded.

 \ms
  In short,  the dark pools are rewarded or penalized by comparing their mutual performances. When the ``attractivity" coefficents $\rho_i$ are not equal, the reasoning is the same but weighted by these attractivities.

\ms
\ni $\rhd$ {\sc Practical implementation.}  One may force the above procedure to stay in the simplex ${\cal P}_{_N}$ by projecting, once   updated, the procedure on   ${\cal P}_{_N}$  each time it exits the simplex. This amounts to replace the possibly negative  $r_i$ by $0$, the $r_i>1$ by $1$ and to renormalize the vector $r$ by dividing it by the sum of its terms. 

Furhermore, to avoid that the algorithm leaves too often the simplex, one may simply normalize the step $\gamma_n$ by considering the predictable step
\[
\tilde \gamma_n =\gamma_n \times \frac{n-1}{V^1+\cdots+V^{n-1}}\approx \frac{\gamma_n}{\E V}.
\] 

\section{The $(IID)$ setting: $a.s$ convergence and $CLT$} \label{Quatre}
\setcounter{equation}{0}
\setcounter{Assumption}{0}
\setcounter{Theorem}{0}
\setcounter{Proposition}{0}
\setcounter{Corollary}{0}
\setcounter{Lemma}{0}
\setcounter{Definition}{0}
\setcounter{Remark}{0}

\begin{Thm} \label{Cvgiid}  Assume that $(V,D)$ satisfy~(\ref{varphiL1}), that $V\!\in L^2(\P)$ and that   Assumption~$({\cal C})$ holds.  Assume furthermore that the distribution of $\frac DV$ satisfies the continuity   Assumption~(\ref{noatom}). Let $\gamma:=(\gamma_n)_{n\ge 1}$ be a step sequence satisfying the usual decreasing step assumption
\[
\sum_{n\ge 1}\gamma_n =+\infty\quad\mbox{ and }\quad  \sum_{n\ge 1}\gamma^2_n <+\infty.
\]
Let $(V^n,D_1^n,\ldots,D_{_N}^n)_{n\ge 1}$ be an i.d.d. sequence defined on a probability space $(\Omega,{\cal A}, \P)$.  Then, there exists an ${\rm argmax}_{{\cal P}_{_N}}\!\Phi$-valued  random variable $r^*$ such that 
\[
r^n \longrightarrow r^*  \quad a.s.
\]
 If the functions $\varphi_i$ satisfy  $({\cal C}_{<})$  then ${\rm argmax}_{{\cal P}_{_N}} \!\Phi\subset int({\cal P}_{_N})$.
\end{Thm}

\medskip 
\noindent {\bf Proof of the theorem.}  In this setting,  the algorithm is (non homogenous) Markov discrete time process with respect to the natural filtration ${\cal F}_n :=\sigma(r^0, (V^k,D^k_1,\ldots,D^k_{_N}), \,1\le k\le n)$ with the following  canonical representation

\begin{eqnarray*}
r^{n+1}&=&r^n+\gamma_{n+1}H(r^n, V^{n+1},D_1^{n+1},\ldots,D^{n+1}_{_N}),\; r^0\!\in {\cal P}_{_N}\\
&=& r^n+\gamma_{n+1}h(r^n)+\gamma_{n+1}\Delta M_{n+1}
\end{eqnarray*}
where, for every $r\!\in {\cal H}_{_N}$, 
\[
h(r) := \E\, H(r,V,D_1,\ldots,D_{_N})= \left(\varphi'_i(r_i)-\frac 1N \sum_{j=1}^N \varphi'_j(r_j)\right)_{1\le i\le N}
\]
is the so-called {\em mean } function of the algorithm, and 
\begin{eqnarray*}
\Delta M_{n+1} & = & H(r^n, V^{n+1},D_1^{n+1},\ldots,D^{n+1}_{_N}) - \E(H(r^n, V^{n+1},D_1^{n+1},\ldots,D^{n+1}_{_N})\,|\,{\cal F}_n)\\
&=&   H(r^n, V^{n+1},D_1^{n+1},\ldots,D^{n+1}_{_N})-h(r^n)
\end{eqnarray*}
since $(V^{n+1}, D^{n+1}_1,\ldots,D^{n+1}_{_N})$ is independent of ${\cal F}_n$.

One derives from Proposition~\ref{MaxiPhi}$(a)$ that the mean function $h$ of the algorithm satisfies $\{h=0\}={\rm argmax}_{{\cal P}_{_N}}$  and that,  for every   $r\!\in {\cal H}_{_N}\!\setminus\!\{h=0\}$ and every $r^*\!\in \{h=0\}$,  
\begin{equation}\label{h}
\langle h(r)\,|\, r-r^*\rangle=\langle h(r)- h(r^*)\,|\, r-r^*\rangle= \sum_{i=1}^N \underbrace{(\varphi'_i(r_i)-\varphi'_i(r^*_i))(r_i-r^*_i)}_{\le 0}<0
\end{equation}
simply because  each function $\varphi'_i$ is non-increasing which implies that each term of the sum is non-positive.  The sum is not zero otherwise  $\varphi'(r_i)=\varphi'(r^*_i)$ as soon as $r_i\neq r^*_i$ which would imply $h(r)=0$.

The random vector $V$ being square integrable, it is clear that $H_i(r,V,D_1,\ldots,D_{_N})$ satisfies the linear growth assumption 
\[
\forall\, i\!\in {\cal I}_{_N},\; \forall\, r\in{\cal H}_{_N},\quad \|H_i(r,V,D_1,\ldots,D_{_N})\|_{_2}\le 2\,(\max_j\rho_j )\|V\|_{_2}(N+|r|)
\]

At this stage one may conclude using a simple variant of the standard Robbins-Monro Theorem (like that established in~\cite{LEPA}): there exists a random variable $r^*$ taking values in $\{h=0\}$ such that $r^n\to r^*$. $\qquad\cqfd$


\subsection{Rate of convergence}

Our aim in this section is to show that the assumptions of the regular Central Limit Theorem ($CLT$) for stochastic approximation procedures are fulfilled. For a precise statement, we refer (among others) to \cite{BEMEPR} (Theorem 13 p.332). For the sake of simplicity, we will assume that the mean function $h$ has a single zero denoted $r^*$.  The following lemma provides a simple criterion to ensure this uniqueness.

\begin{Lem} \label{Unic} Assume that all the functions $\varphi_i$, $i\!\in {\cal I}_{_N}$,  are decreasing (strictly). Then
\[
\{h=0\}= {\rm argmax}_{{\cal P}_{_N}}\!\Phi = r^*\!\in int({\cal P}_{_N}).
\] 
\end{Lem}

\noindent {\bf Proof.} In particular $({\cal C}_<)$ is satisfied so that ${\rm argmax}_{{\cal P}_{_N}}\!\Phi \subset r^*\!\in int({\cal P}_{_N})$. If $r$, $r'\!\in \{h=0\}$, $r\neq r'$, it follows from~(\ref{h}) that $\varphi'_i(r_i)=\varphi'_i(r'_i)$ for some index $i$ such that $r_i\neq r'_i$. $\cqfd$ 

\bigskip
The second ingredient needed to establish a $CLT$ will be the Hessian of function $\Phi$. To ensure its existence we will make one further assumption on a generic random couple $(V,D)$, keeping in mind that $\P(D>0)>0$, but that $\P(D=0)$ may possibly be positive too.  Namely,  assume  that the   distribution function of $(V,D)$ {\em given } $\{D>0\}$ is absolutely continuous with a probability density $f$ defined on $(0,+\infty)^2$. Furthermore we make the following assmptions on $f$:

\begin{equation}\label{densitecond}
\left\{\begin{array}{ll}
(i)&\mbox{for every $v>0$, } u\mapsto f(v,u) \hbox{ is continuous and positive on $(0,\infty)$,}\\\\
(ii)&\displaystyle \forall\, \varepsilon\!\in (0,1),\quad  \sup_{\varepsilon V \le u\le V/\varepsilon} \hskip -0.25 cm f(V,u)V^2\!\in L^1(\P).
\end{array}\right.
\end{equation}

Note that $(ii)$ is clearly always satisfied when $V\!\in L^2(\P)$ and $f$ is bounded. The conditional distribution function of $D$ given $\{D>0\}$ and $V$ is given by 
$$
F_{_D}(u\,|\,V=v,D>0):=\P(D\le u\,|\, V=v, D>0)=\int_0^uf(v,u')du' , \;u> 0, \;v>0,
$$
  \begin{Lem} \label{Lem:2}$(a)$ Assume  $(V,D)$ satisfies the above assumption~(\ref{densitecond}). 
Then  the mean execution function $\varphi(u): = \rho\,\E (\min (uV,D))$ is concave,  twice differentiable on $\R_+$ and 
for every $u>0$, 
\[
\varphi''(u) =-\rho\,\E\left(V^2\mbox{{\bf 1}}_{\{D>0\}}f(V,uV)\right)<0.
\] 

\medskip
\noindent $(b)$ If  $(V,D_i)$ satisfies the above assumption~(\ref{densitecond}) for every~$i\!\in {\cal I}_{_N}$, then the   function $\tilde h$ defined on  $\R_+^N$  by $\tilde h(u_1,\ldots,u_{_N})=\left(\varphi'_i(u_i)-\frac 1N  \sum_{1\le j\le N} \varphi'_j(u_j)\right)_{1\le i\le N}$  is differentiable on $(0,\infty)^N$ and admits a  continuous extension on $\R_+^N$ given by 
\[
D\tilde h(u)= -\frac 1N \Big[-a_j (u_j)+N a_i(u_i)\,\delta_{ij}\Big]_{1\le i,j\le N}\quad \mbox{ with } \quad a_i(u) =
-\varphi_i''(u)>0.
\]

\noindent $(c)$  Let $A:=[-a_j+N a_i\delta_{ij}]_{1\le i,j\le N}$, $a_1,\ldots,a_{_N}>0$ and let  $\underline a =\min_i a_i$.  Its kernel ${\rm Ker}(A)$ is one dimensional, $A(\R^N)= \mbox{\bf 1}^\perp$ and $A_{|\mbox{\bf 1}^\perp}$ is bijective. Every non-zero eigenvalue $\lambda$ (with eigenspace $E_{\lambda}$) satisfies 
\[
\Re(\lambda)\ge N\times\underline a\quad \mbox{ and }\quad E_{\lambda} \subset \mbox{\bf 1}^\perp.
\]
 
\end{Lem}

\noindent {\bf Proof.} $(a)$ is a straightforward application of the Lebesgue differentiation Theorem for expectation.   

\smallskip
\noindent $(b)$ is a consequence of $(a)$. 

\smallskip
\noindent $(c)$  The transpose $A^t$ of $A$ has a strict dominating diagonal structure $i.e.$ $A^t_{ii}>0$, $A^t_{ij}<0$, $i\neq j$ and $\sum_jA^t_{ij} =0$ for every $i$. Consequently, it follows from Gershgorin's Lemma (see~\cite{GER}) that  $0$ is an eigenvalue of order $1$ of $A^t$ (with $\mbox{\bf 1}$ as an eigenvector  and that all other eigenvalues have (strictly) positive real parts). Consequently ${\rm Ker}(A)$ is one dimensional.  The fact that $A(\R^N)\subset \mbox{\bf 1}^\perp$ is obvious so that this inclusion  holds as an equality by the dimension formula. Hence all the eigenvectors not in   ${\rm Ker}(A)$ are in $\mbox{\bf 1}^\perp$. Set $\tilde a_i-a_i-\underline a\ge 0$, $i=1,\ldots,N$. Then $\tilde A^t$ has a dominating diagonal structure so that all its eigenvalues have non-negative real parts. Now if $\lambda $ is an eigenvalue of $A$, it is obvious that $\lambda-N\underline a$ is an eigenvalue of $\tilde A$.  Consequently $\Re (\lambda)\ge N\underline a$. ~$\cqfd$

\begin{Thm} Assume that  the assumptions of Theorem~\ref{Cvgiid} holds and that ${\rm argmax}\,\Phi$ is reduced to a single point $r^*\!\in{\cal P}_{_N}$ so that $r^n \to r^*$ $\P$-$a.s.$ as $n\to \infty$.    Furthermore, suppose that  Assumption~(\ref{densitecond}) holds  for every $(V,D_i)$, $i\!\in {\cal I}_{_N}$ and that  $V\!\in L^{2+\delta}(\P)$, $\delta>0$. 
Set
\[
\gamma_n =\frac{c}{n},\; n\ge 1\; \mbox{ with }\;c>\frac{1}{2\Re e (\lambda_{\min})} 
\]
where $\lambda_{\min}$ denotes the eigenvalue of $A^\infty:=-Dh(r^*)_{|\mbox{\bf 1}^\perp}$ with the lowest real part. Then

\[
\frac{r^n -r^{*}}{\sqrt{\gamma_n} } \stackrel{\cal L}{\longrightarrow}{\cal N}(0;\Sigma^\infty)
\]
where the asymptotic covariance matrix $\Sigma^\infty$  is given by 
\[
\Sigma^{\infty}= \int_0^{\infty}e^{u  (A^\infty-\frac{Id}{2c})}C^\infty e^{u(A^\infty-\frac{Id}{2c})^t}du
\]
where 

\[
C^\infty = \E \, \left (H(r^*,V,D_1,\ldots,D_{_N})H(r^*,V,D_1,\ldots,D_{_N})^t\right)_{| \mbox{\bf 1}^\perp}
\]
and $(A^\infty-\frac{Id}{2c})^t$ stands for  the {\em transpose  operator} of $A^\infty-\frac{Id}{2c}\!\in {\cal L}(\mbox{\bf 1}^\perp)$.
\end{Thm}

\noindent {\bf Remark.} The above claim is consistent since $u\mapsto H(r,v,\delta_1,\ldots,\delta_{_N})^t\,u $ preserves $\mbox{\bf 1}^\perp$.

\bigskip
\noindent {\bf Proof.} First note that,  since $r^*\!\in int({\cal P}_{_N})$, the above Lemma~\ref{Lem:2}$(b)$ shows that (still making the confusion between the linear operator $Dh(r^*)$ and its matrix representation in the canonical basis) 
\[
Dh(r^*)= -\frac 1N \left[-a_j(r_j^*) +N a_i(r_i^*)\,\delta_{ij}\right]_{1\le i,j\le N}\quad \mbox{ with } a_i (r)=\rho_i  \E (V^2\mbox{{\bf 1}}_{\{D_i>0\}}f(V,rV))>0
\]
 Then, Lemma~\ref{Lem:2} $(c)$ implies that  $-Dh(r^*)_{|\mbox{\bf 1}^\perp}$ has eigenvalues with positive real parts, all lower bounded by $\min_i a_i(r_j^*)>0$. 
 
 At this stage, one can apply the $CLT$ for stochastic algorithms defined on $\mbox{\bf 1}^\perp$ (see $e.g.$~\cite{BEMEPR}, p.341). ~$\cqfd$

%

\section{The $(ERG)$ setting: convergence}\label{Cinq}
\setcounter{equation}{0}
\setcounter{Assumption}{0}
\setcounter{Theorem}{0}
\setcounter{Proposition}{0}
\setcounter{Corollary}{0}
\setcounter{Lemma}{0}
\setcounter{Definition}{0}
\setcounter{Remark}{0}

  For the sake of simplicity, although it is not really necessary, we will assume throughout  this section  that  
  \[
  {\rm argmax}_{{\cal P}_{_N}}\, \Phi= \{r^*\}\subset int({\cal P}_{_N})
  \]
possibly  because  all the execution functions $\varphi_i$ are decreasing so that, following the former Lemma~\ref{Unic}.

    


So we assume that 
the sequence $(V^n,D^n_i, i=1,\ldots,N)_{n\ge 1}$ satisfies $(ERG)_{\nu}$ with a limiting distribution 
%
$\nu$ such that, for every $i\!\in {\cal I}_{_N}$, its marginal  $\nu_i={\cal L}(V,D_i)$ satisfies the consistency assumption~(\ref{varphiL1}) and  the continuity assumption~(\ref{noatom}). We will also need to make a specific assumption: there exists $\e_0>0$ such that 
\begin{equation}\label{dispersion}
\left\{\begin{array}{ll}
(i) &\P(V\ge \e_0)>0\\
(ii)& {\rm supp}\!\left(\!{\cal L}\!\left(\!\frac{D_i}{V}, i=1,\ldots,N\,|\, \{V\ge \e_0\}\right)\!\right)\mbox{ is a neighbourhood of ${\cal P}_{_N}$ in $\R_+^N$}.
\end{array}\right.
\end{equation}
This assumption means that {\em all  allocations across the pools lying  in the neihbourhood of } ${\cal P}_{_N}$ can be executed.

On the other hand, it follows from~$(ERG)_{\nu}$ and some standard weak 
convergence arguments that  
\[
\forall\, i\!\in {\cal I}_{_N},\; \forall\,u\!\in \R_+,\quad   \frac{1}{n} \sum_{k=1}^n V^k\mbox{\bf 1}_{\{uV^k\le D_i^k\}}-\E(V \mbox{\bf 1}_{\{u V\le D_i\}})\stackrel{a.s.\& L^2}{\longrightarrow} 0\quad \mbox{ as } n\to \infty,
\]
since the (non-negative) functions $f_u(v,\delta):= v\mbox{\bf 1}_{\{uv\le \delta\}}$, $u>0$, are  $\P_{(V,D_i)}$-$a.s.$ continuous and $O(v)$ as $v\to + \infty$ by~(\ref{noatom}). Moreover this $a.s.$ convergence holds uniformly on compact sets with respect to $u$ since $u\mapsto \E V\mbox{\bf 1}_{\{uV\le D_i\}}$ is continuous, still owing to~(\ref{noatom}). Our specific assumption is to require a rate in the above $a.s.$ and $L^2(\P)$-convergence. Namely, we assume that there exists an exponent $\alpha_i\!\in (0,1]$ such that
\begin{equation}\label{HypoErgo}
\forall\,u\!\in \R_+,\quad \frac{1}{n} \sum_{k=1}^n V^k  \mbox{\bf 1}_{\{uV^k < D_i^k\}}-\E(V \mbox{\bf 1}_{\{uV < D_i\}})= O(n^{-\alpha_i})\quad a.s. \ \mbox{and in} \ L^2(\P).
\end{equation}

This assumption $e.g.$ from the more general assumption that, for every $i\in{\cal I}_N$, the marginal $\nu_i={\cal L}(V,D_i)$ satisfies (\ref{noatom}) and
$$(V^n,D_i^n) \mbox{ is $\nu_i$-averaging at rate $\alpha_i$}$$
on a subspace ${\cal V}_{\alpha_i,2}$ containing all the functions $f_u$. 

Note that, when the sequence $(V^n,D_i^n, i=1,\ldots,N)_{n\ge 1}$ is i.i.d. with distribution $\nu$ then elementary martingale arguments show that the whole sequence is $\nu$-averaging at  rate $\frac 12 -\eta$ for every $\eta\!\in (0,1/2)$ on ${\cal V}_{\frac 12 -\eta,2}=L^2(\nu)$ (and all $f_u\in L^2(\nu)$, $u>0$, since $V\in L^2(\P)$). So, the theorem below almost embodies the $a.s.$ convergence theorem established in the $(IID)$ setting (except for the integrability assumption on $V$).

Now we are in position to state the main  convergence result of this section. We rely on the extension of Robbins-Siegmund Lemma proposed in~\cite{LARPAG}. For the reader's convenience it is recalled in the Appendix. 

\begin{Thm}
Let  $(V^n, D_1^n,\ldots,D_N^n)_{n\ge 0}$ be a  sequence of input satisfying $(ERG)_{\nu}$ and such that,  for every $i\!\in {\cal I}_{_N}$, the marginal distribution  $\nu_i ={\cal L}(V,D_i)$ satisfies the consistency assumptions~(\ref{varphiL1}) and~(\ref{noatom}).
Suppose furthermore that,   
the sequence $(V^n,D_i^n)_{n\ge 1}$ satisfies the rate assumption~(\ref{HypoErgo}). 
 If the step sequence $(\gamma_n)_{n\ge 1} $ satisfies 
\[
\sum_{n\geq1}\gamma_n=+\infty, \quad
\gamma_n = o(n^{\underline \alpha-1})\quad \mbox{ and }\quad \sum_{n\ge 1} n^{1-\underline \alpha}\max(\gamma_n^2,|\gamma_n-\gamma_{n+1}|)<\infty 
\]
where $\underline \alpha :=\min _{i\in {\cal I}_{_N}} \alpha_i\!\in (0,1]$, then the algorithm defined by~(\ref{LAlgo}) 
 $a.s.$ converges towards $r^*={\rm argmax}_{{\cal P}_{_N}}\,\Phi$.
\end{Thm}

\noindent {\sc technical comment.} The above condition on the step sequence $(\gamma_n)_{n\ge1}$ is satisfied as soon as $\gamma_n= \frac{c}{n^{\beta}}$ with $\beta\!\in (1-\underline \alpha,1]$.

\bigskip
\noindent {\bf Proof.} {\sc Step~1.} First, we aim at applying the extended Robbins-Siegmund Lemma   established in~\cite{LARPAG} (see Appendix, Theorem~A.1 for its statement)  for stochastic algorithms with $\nu$-averaging   inputs dynamics in presence of a Lyapunov function. We will consider the case  $p=2$ and $\beta \!\in (0,\underline \alpha]$. We set $G=-H$ and $\Delta M^n \equiv 0$ and we consider the input $Y^n= (V^{n+1}, D_1^{n+1},\ldots,D_N^{n+1})$, $n\ge 0$. Let $L(r)= \frac 12 |r-r^*|^2$ be our candidate as a  Lyapunov function.

First note that it follows from~(\ref{LAlgo})  that  the function $H$ satisfies the growth assumption~(\ref{Lyapunov}) since
$$
\forall r\in{\cal H}_n, \ \forall y\in {\R}^{N+1}, \quad \left|H(r,y)\right|\leq C_H g(y)(1+|r|)
$$
where $C_H>0$ and $g(v,\delta_1,\ldots,\delta_{_N})=v$.

In view of the ergodic assumption~(\ref{HypoErgo}) and the fact that $r^*$ lies in ${\cal P}_{_N}$, it is clear from its definition  that $H(r^*,.)\!\in {\cal V}_{\beta,2}$ for every $\beta\!\in (0, \underline \alpha]$.  
 
At this stage it remains to check the ``weak  local Lyapunov" assumption~(\ref{lmr}) for $G=-H$. This fact is obvious since, for every $ r\!\in {\cal H}_{_N}$ and every input $y=(v,\delta_1,\ldots,\delta_N)\!\in (0,+\infty)\times \R^N$, 
\[
 \langle H(r,y)-H(r^*,y) |r-r^*\rangle=\sum_{i=1}^N (\widetilde{H}_i(r_i,v,\delta_i)-\widetilde{H}_i(r^*_i,v,\delta_i))(r_i-r^*_i) \le 0
\]
where 
\begin{equation}\label{Phi-i}
\widetilde{H}_i(u,v,\delta_i) = \rho_iv\left(\mbox{\bf 1}_{\{uv\le \delta_i\}}\mbox{\bf 1}_{[0,1]}(u) +(1-u)\mbox{\bf 1}_{\delta_i>0, u<0}+\frac {1}{u}\mbox{\bf 1}_{v\le \delta_i,u>1}\right),\; i\!\in {\cal I}_{_N}
\end{equation}
is clearly non-increasing with respect to $u$. 

At this stage, using that $\sup_{n\ge 1} \E (V^n)^2 <+\infty$, we can apply our extended Robbins-Siegmund lemma also that 
\begin{equation}\label{RZ}
|r^n-r^*|\stackrel{a.s.}{\longrightarrow} L_{\infty}<+\infty \; a.s.\;\mbox{ and }\quad \sum_{n\ge 1}\g_n\langle r^n-r^*\,|\, G(r^n,Y^{n})-G(r^*,Y^n)\rangle<+\infty \; a.s.
\end{equation}

\ss
\ni {\sc Step~2.} At this stage it suffices to show that $r^*$ is $a.s.$ a limiting point of $(r^n)_{n\ge 0}$ since $|r^n-r^*|$ converges to $L_{\infty}<+\infty$ $a.s.$

Let $\eta$ denote a  positive real number such that, for every $i\!\in {\cal I}_{_N}$, $[r^*_i-\eta, r^*_i+\eta]\subset (0,1)$. One derives from~(\ref{Phi-i})  and the monotonicity of $\widetilde{H}_i(u,v,\delta)$ in $u\!\in \R$ that for every $i\!\in {\cal I}_{_N}$ and every $r\!\in {\cal H}_{_N}$, 
\[
(\widetilde{H}_i(r_i,v,\delta_i)-\widetilde{H}_i(r^*_i,v,\delta_i))(r^*_i-r_i) \ge \rho_iv\eta\mbox{\bf 1}_{\{r_i>r^*_i+\eta\}}\mbox{\bf 1}_{\{\delta/v\in J_{\eta}\}}
\]

\noindent where $J_{\eta}= (r^*_i,r^*_i+\eta)$. As a consequence
\[
\langle G(r,y)-G(r^*,y)\,|\, r-r^*\rangle\ge \e_0\underline \rho\,\eta \mbox{\bf 1}_{\{v\ge \e_0\}}\mbox{\bf 1}_{y\in O_{\eta}}\!\sum_{i\in {\cal I}_{_N}}\!\!\mbox{\bf 1}_{r_i>r^*_i+\eta}.
\]
where $\underline \rho =\min_i\rho_i$ and  the open set $O_{\eta}$ is defined by 
$$
O_{\eta}=\left \{y=(v,\delta_1,\ldots,\delta_N)\!\in (\e_0,+\infty)\times \R_+^N\,\hbox{ s.t. }\, \frac{\delta_i}{v}\in J_{\eta}, \, i\in {\cal I}_{_N}\right\}.
$$

 Now, one derives from~(\ref{RZ})  that 
 \[
 \sum_n \g_n\mbox{\bf 1}_{O_{\eta}}(Y^n)\sum_{i\in {\cal I}_{_N}}\mbox{\bf 1}_{r^n_i>r^*_i+ \eta} <+\infty\quad a.s.
 \]

Now Assumption~(\ref{dispersion}) implies that $\nu(O_{\eta})>0$. Furthermore $\nu(\partial O_{\eta})=0$ owing to the continuity assumption so that $(ERG)_{\nu}$ implies
\[
\frac 1n \sum_{k=1}^n \mbox{\bf 1}_{O_{\eta}}(Y^k)\longrightarrow \nu(O_{\eta})>0\quad a.s.
\] 
An Abel transform and the facts  that the sequence $\g_n$ is non-increasing  and $\sum_{n\ge 1}\g_n =+\infty$ classically implies that
\[
\sum_{n\ge1} \g_n ( \mbox{\bf 1}_{O_{\eta}}(Y^k)-\nu(O_{\eta}))\quad\mbox{$a.s.$ converge}
\]
so that
\[
\sum_n\g_n \mbox{\bf 1}_{O_{\eta}}(Y^n) =+\infty \quad a.s.
\]
In turn, this  implies that
 \[
\liminf_n \sum_{i\in {\cal I}_{_N}}\mbox{\bf 1}_{\{r^n_i>r^*_i+ \eta\}} =0\quad  a.s.
 \]
This holds of course for a sequence of real numbers $\eta^\ell$ decreasing to $0$.

Let ${\cal R}_\infty$ be the set  of limiting values of the sequence $(r^n)_{n\ge0}$. It is is $a.s.$ non-empty since the sequence $(r^n)_{n\ge 0}$ is bounded. Then ${\cal R}_\infty$ is $a.s.$  compact   and it follows from what precedes that  ${\cal R}_\infty\cap \prod_{1\le i\le N}(-\infty,r^*_i+ \eta^\ell ]\neq \emptyset$ (and is compact). Hence a decreasing intersection of non-empty compact sets being a (non-empty) compact set ${\cal R}_\infty\cap \prod_{1\le i\le N}(-\infty,r^*_i ]\neq \emptyset$. On the other hand ${\cal R}_{\infty} \subset {\cal H}_{_N}$ since the algorithm is ${\cal H}_{_N}$-valued. But  $ \prod_{1\le i\le N}(-\infty,r^*_i ]\cap {\cal H}_{_N}= \{r^*\}$. Consequently $r^*$ is a limiting point of the algorithm which implies that it is its true $a.s.$ limit.~$\cqfd$

\bigskip
\noindent {\sc Application to $\alpha$-mixing stationary data.}  If $(V^n, D^n_i, i=1,\ldots,N)_{n\ge 1}$ is a stationary $\alpha$-mixing  sequence  which mixing coefficients $(\alpha_n)_{n\ge1}$ satisfy Ibragimov's condition for some $\delta>0$:
\[
\sum_{n\ge1} \alpha_n^{\frac{2}{2+\delta}}<+\infty 
\]
(which is satisfied in case of geometric $\alpha$-mixing) then the sequence $(V^n,D^n_i, i=1,\ldots,N)_{n\ge 1}$  is $\nu$-averaging where $\nu$ is the stationary marginal distribution  of the sequence (supposed to satisfy~(\ref{varphiL1}) and~(\ref{noatom})) at rate $\beta$ for every $\beta\!\in (0,1/2)$. To be precise,    $L^2(\nu)\subset {\cal V}_{0^+,2} $ and
\[
L^{2+\delta}(\nu)\subset \bigcap_{0<\beta<\frac12}{\cal V}_{\beta,2}.
\]
In particular, all the functions $f_u(v,\delta):= v\mbox{\bf 1}_{\{uv\le \delta\}}$, $u\ge 0$, lie in every ${\cal V}_{\beta,2}$, $0<\beta<\frac 12$, so that the rate condition~(\ref{HypoErgo}) is satisfied.

\medskip
As concerns the stationary assumption on  the input data sequence, it  can be considered as  realistic if one think of execution objectives given  on a daily basis.

\bigskip
%
%
%
%
\noindent{\sc Example:}  An exponential discrete time Ornstein-Ulhenbeck model for  $(V^n,D^n_1,\ldots,D^n_{_N})$.  

\[
V^{n} = v^0e^{X_0^n} , \qquad D_i^n =d_i^0e^{X_i^n},\; i=1,\ldots,N,\; n\ge 1,
\]
where $v^0$, $d_1^0$, \dots, $d_{_N}^0$ are positive real numbers and the sequence $(X^n)_{n\ge 1}$ satisfies the linear auto-regressive dynamics
\[
  X^{n+1} = m+A  X^{n}+ B \,\Xi^{n+1}, n\ge 1,
\]
with $m\in{\R}^{N+1}$, $A\!\in {\cal M}(N+1,N+1,\R)$, $|\!|\!|A|\!|\!|<1$,  $B\!\in {\cal M}(N+1,M)$  with rank$(B)=N+1$ ($\le M$)   and $(\Xi^n)_{n\ge1}$ is an i.i.d. sequence of ${\cal N}(0; Id_{M})$-distributed random variables.  We assume that the sequence is stationary $i.e.$ that the distribution of $X^1$ is the (Gaussian) invariant distribution  (with covariance matrix $C$ solution to the Lyapunov equation $C-ACA^t=BB^t$ where $^t$ stands for transpose). Then (see~\cite{DOU}, p. 99), the sequence $(X^n)_{n\ge 0}$ is geometrically $\alpha$-mixing  and subsequently so is $(V^{n},D_1^n,\ldots,D_{_N}^n)_{n\ge 1}$ (with respect to its natural filtration). Furthermore, it is clear that its distribution $\nu$ satisfies the dispersion assumption~(\ref{dispersion}) the process $(X^1_0,X^1_i-X^1_0, i=1,\ldots,N)$ is a non-degenerate Gaussian distribution over $\R^N$ since $B$ has full rank $N+1$.

%

\section{An alternative procedure  based on a  reinforcement principle.}\label{Six}
\setcounter{equation}{0}
\setcounter{Assumption}{0}
\setcounter{Theorem}{0}
\setcounter{Proposition}{0}
\setcounter{Corollary}{0}
\setcounter{Lemma}{0}
\setcounter{Definition}{0}
\setcounter{Remark}{0}

Recently, inspired by  the discussion developed by Almgren and Harts  in~\cite{almsor08}   about liquidity estimation, Berenstein and Lehalle devised a  ``smart routing" recursive procedure of requests to be executed by a pool of $N$   dark pools (see~\cite{BELE}). This procedure is not  based on the opimization
of a potential function but  on a intuitive reinforcement mechanism. Let $I_i^n$ be the profit induced by the execution of the order sent  to dark pool $i$ at time $n$.  The proportion $r^n_i$ of the global order $V^{n+1}$ to be sent to dark pool $i$ for execution at time $n+1$ is defined as proportional to this profit $i.e.$ by
\[
\forall\, i\!\in {\cal I}_{_N},\qquad r^n_i:=\frac{I^n_i}{\sum_jI^n_j}.
\]
The updating of the random vector $I^n$ is as follows
 \[
\forall\, n\ge 0,\;\forall\, i\!\in {\cal I}_{_N},\quad I^{n+1}_i=I^n_i+\rho_i\,\min\left(r^n_iV^{n+1},D^{n+1}_i\right), \quad I_i^0=0.
\]
The first equation models the idea of  ``reinforcement" since the proportion of orders  sent for execution to  dark pool $i$ is  proportional to the historical performances of this dark pool since the beginning of the procedure. 

The second equation describes in a standard way --~like in the optimization algorithm~-- the way dark pools execute orders.

Elementary computations show that the algorithm can be written directly in a recursive way in terms of a new vector valued variable 
\[
X^n =\frac{I^n}{n},\qquad n\ge 1,
\]
since 
\[
X_i^{n+1}=X_i^n -\frac{1}{n+1}\left(X^n_i-\rho_i\,\min \left(r^n_iV^{n+1},D^{n+1}_i\right)\right),\quad i\!\in {\cal I}_{_N}.
\]
This is a standard form a stochastic algorithm (with step $\gamma_n=\frac 1n$). 

Furthermore, note that setting $\underline{\rho}:=\min_i \rho_i $, 
\[
\sum_{ i \in {\cal I}_{_N}}I^n_i \ge \underline{\rho}   \min \left(\frac{1}{N} V^{n}, \min_{i\in {\cal I}_{_N}}D^{n}_i\right) 
\]
since $r^n_i\ge \frac 1N$ for at least  one dark pool $i\!\in {\cal I}_{_N}$. Consequently, as soon as  the sequence $(V^n,D_1^n,\ldots,D^n_{_N})$ is stationary and ergodic
 \[
\liminf_n \sum_{i\in {\cal I}_{_N}} X^n_i \ge\underline{\rho}\lim_n \frac 1n \sum_{k=1}^n \min \left(\frac{1}{N} V^{k}, \min_{i\in {\cal I}_{_N}}D^{k}_i\right)=  \underline{\rho}\, \E \min  \left(\frac{1}{N} V, \min_{i\in {\cal I}_{_N}}D_i\right)   \;a.s.
\]
So if we make the natural assumption that 
\[
\E \min\left(\frac{1}{N} V, \min_{i\in {\cal I}_{_N}}D_i\right) >0
\]
then, $a.s.$,   the algorithm $X^n$ cannot converge to $0$.

\medskip
If we make the additional assumption that the sequence $(V^n,D_1^n,\ldots,D^n_{_N})$ is $i.i.d.$ then the algorithm is a discrete time (non homogenous)  ${\cal F}_n$-Markov process with respect to the filtration ${\cal F}_n=\sigma(V^k,D_1^k,\ldots,D_{_N}^k,k=1,\ldots,n)$, $n\ge0$, so that it admits the canonical representation

\[
X_i^{n+1}=X_i^n -\gamma_{n+1}\left(X_i^n-\varphi_i(r_i^n)\right)+\gamma_{n+1}\Delta M_i^{n+1 }\quad i\!\in {\cal I}_{_N},\quad n\ge 0,
\]
where  $\gamma_n=\frac 1n$ and 

\[
\Delta M_i^n =\rho_i \min \left(r_i^{n-1}V^{n},D^{n}_i\right)-\varphi_i(r^{n-1}_i),\quad i\!\in {\cal I}_{_N},\quad n\ge 1, 
\]
is an ${\cal F}_n$-martingale increment. Furthermore it is $L^2$-bounded as soon as $V\!\in L^2$.

In fact the specific difficulties induced by this algorithm are more in relation with its mean function 
\begin{equation}\label{h2}
h:x\longmapsto \left(x_i-\varphi_i\left(\frac{x_i}{\sum_j x_j}\right)\right)_{1\le i\le N}
\end{equation}
  than with the martingale ``disturbance term"  $\gamma_{n+1}\Delta M^{n+1}$. 
Our first task will be to prove under natural assumptions the existence of a non degenerate equilibrium point. Then we will show  why this induces the existence of many parasitic equilibrium points.

 \subsection{Existence of an equilibrium}
In this section, we will need to introduce a new function  associated to a generic order $V$ and a generic dark pool with characteristics $(\rho,D)$.
\begin{equation}\label{psi}
\psi(u) := \frac{\varphi(u)}{u},\quad u>0,\qquad \psi(0) = \varphi'(0)=\rho\, \E\, V\mbox{\bf 1}_{\{D>0\}}.
\end{equation}
 If  Assumption~(\ref{varphiL1}) holds then $\psi(0)<+\infty$ and $\psi$ is continuous at $0$. It follows from the concavity of $\varphi$ and $\varphi(0)=0$ that $\psi $ is non-increasing.   It is continuous as soon as $\varphi$ is $e.g.$ if Assumption~(\ref{noatom}) holds true.

\begin{Pro} \label{ExistEquil} Let $N\ge1$. Assume that Assumption~(\ref{varphiL1}) holds for every couple $(V,D_i)$, $i\!\in {\cal I}_{_N}$. 

\medskip
\noindent $(a)$ There exists a   $x^*\!\in\R_+^N$ such that
\begin{equation}\label{Equil}
\sum_{i\in {\cal I}_{_N}}x^*_i>0\quad \mbox{ and }\quad \varphi_i\left(\frac{x_i^*}{\sum_{j\in {\cal I}_{_N}}x^*_j}\right)=x^*_i,\quad i\!\in {\cal I}_{_N}.
\end{equation}

\medskip
\noindent $(b)$ Let $\psi_i$ be the functions associated to   dark pool $i\!\in {\cal I}_{_N}$ by~(\ref{psi}). Assume  that for every $i\!\in {\cal I}_{_N}$, $\psi_i$ is (continuous and) decreasing on $[0,\infty)$ and that
\begin{equation}\label{Equil:2}
\sum_{i\in {\cal I}_{_N} } \psi_i^{-1}(\min_{i\in {\cal I}_{_N}}\varphi'_i(0))<1.
\end{equation}
Then there exists $x^*\!\in int({\cal P}_{_N})$ satisfying~(\ref{Equil}). 

\end{Pro}

\noindent {\bf Proof.}  $(a)$  We define for every $r=(r_1,\ldots,r_{_N})\!\in {\cal P}_{_N}$  
\[
 \Psi(r):= \left(\frac{\varphi_i(r_i)}{\sum_{j\in {\cal I}_{_N}}\varphi_j(r_j)}\right)_{i\in {\cal I}_{_N}}.
\]
This function maps the compact convex set ${\cal P}_{_N}$ into itself. Furthermore it is continuous since, on the one hand, for every $i\!\in {\cal I}_{_N}$, $\varphi_i$ is continuous owing to the fact that $(V,D_i)$ satisfies~$(\ref{varphiL1})$ and, on the other hand,  
\[
\sum_{j\in {\cal I}_{_N}}\varphi_j(r_j)\ge \min_{j\in {\cal I}_{_N}}\varphi_j\left(\frac 1N\right)>0.
\]
Indeed, for every $i\!\in {\cal I}_{_N}$, 
\[
\varphi_j\left(\frac 1N\right)\ge \frac 1N \E \, \min(V,D_i)>0
\]
since $V>0$ $\P$-$a.s.$ and $\P(D_j=0)<1$. Then it follows from the Brouwer Theorem that the function $\Psi$ has a fixed point $r^*$. Set for every $ i\!\in {\cal I}_{_N}$, 
\[
x^*_i= r^*_i\sum_{j\in {\cal I}_{_N}}\varphi_j(r_j).
\]
It follows immediately from this definition that
\[
\forall\, i\!\in {\cal I}_{_N},\quad x^*_i = \varphi_i(r^*_i)
\] 
which in turn implies that $\sum_{j\in {\cal I}_{_N}}\varphi_j(r^*_j)=\sum_{j\in {\cal I}_{_N}}x^*_j$ so that $r^*_i=\frac{x^*_i}{\sum_{j\in {\cal I}_{_N}}x^*_j}$, $i\!\in {\cal I}_{_N}$.

\medskip
\noindent $(b)$ For every $i\!\in {\cal I}_{_N}$ we consider the  inverse of $\psi_i$ defined on  the interval $(0,\varphi'_i(0)]$. 
This function is decreasing continuous and $\lim_{v\to 0} \psi_i^{-1}(v)=+\infty$. Then, let $\Theta$ be the continuous  function  defined by
\[
\forall\, \theta \!\in (0, \min_{i\in {\cal I}_{_N}}\varphi'_i(0)],\quad \Theta(\theta)= \sum_{i\in {\cal I}_{_N}}\psi_i^{-1}(\theta).
\]
We know that $\lim_{\theta\to 0} \Theta(\theta)=+\infty$ and we derive from Assumption~(\ref{Equil:2}) that $\Theta(\min_{i\in {\cal I}_{_N}}\varphi'_i(0))\le 1$. So, owing to the (strict) monotonicity of $\theta$, there exists $\theta^*\!\in (0, \min_{\in {\cal I}_{_N}}\varphi'_i(0))$ such that $\Theta(\theta^*)=1$. Set
\[
r^*_i= \psi_i^{-1}(\theta^*), \quad i\!\in {\cal I}_{_N}.
\]
Then $r^*:=(r^*_1,\ldots,r^*_{_N})\!\in int({\cal P}_{N})$   since $\sum_i r^*_i=1$ by definition of $\theta^*$.  If $r^*_{i_0}=0$, then $\theta^*= \psi_{i_0}(0)=\min_{i\in {\cal I}_{_N}}\varphi'_i(0)$ which is impossible. $\cqfd$

\begin{Cor} Assume that all the functions $\psi_i$ are continuous and decreasing. If furthermore, the rebate coefficients $\rho_i$ are equal (to $1$) and if $\P(D_i=0)=0$ for every $i\!\in {\cal I}_{_N}$ then 
there exists an equilibrium point lying in $int({\cal P}_{_N})$.
\end{Cor}

\medskip
\noindent {\bf Proof.} Under the above assumptions $\varphi'_i(0) = \E \, V>0$.  Consequently 
$$
\psi_i^{-1}(\min_{i\in {\cal I}_{_N}}\varphi'_i(0))= \psi^{-1}_i(\psi_i(0))=0<1. \cqfd
$$

\smallskip
\noindent{\sc Comments.} Unfortunately there is no hope to prove that all the equilibrium points lie in the interior of ${\cal P}_{_N}$ since one may always adopt an execution strategy which boycotts a given dark pool or, more generally, $N_0$ dark pools. So it seems hopeless to get uniqueness of the equilibrium point. To be more precise, under the assumptions of claim~$(b)$ of the above Proposition~\ref{ExistEquil}, there exists at least one strategy involving a subset of $N-N_0$ dark pools $N_0=0,\ldots,N-1$ (one dark pool is needed at least). Elementary combinatorial arguments  show that there are {\em at least} $2^N-1$ equilibrium points. 

So, from a theoretical point of view,  we are facing a situation where there may be  many parasitic equilibrium points, some of them being clearly parasitic. However it is quite difficult to decide {\em a priori}, even if we make the unrealistic assumption that we know all the involved distributions, which equilibrium points are parasitic.    

This is  a typical situation encountered when dealing with procedures devised from a reinforcement principle. 

However, one may reasonably hope that some of them are so-called ``traps", that means equilibrium points which are repulsive at least in one noisy direction so that the algorithm escapes from it.
Another feature described below suggests that a theoretical study of the convergence behaviour of this procedure would need a specific extra work.

\bigskip The next natural question is to wonder whether an equilibrium $x^*$  of the algorithm -- namely a zero of $h$~-- is  (at least) a target for the algorithm $i.e.$ is attractive for the companion $ODE$, $\dot x =   - h(x)$.

\begin{Pro} An equilibrium $x^*$ satisfying~(\ref{Equil}) is locally uniformly attractive  as soon as 
\[
\sum_{j\in {\cal I}_{_N}}\frac{x^*_j}{(\bar x^{*})^2}\varphi_j'\left(\frac{x^*_j}{\bar x^*}\right)<1-\frac{1}{\bar x^*}\max_{i\in {\cal I}_{_N}}\varphi'\left(\frac{x_i^*}{\bar x^*}\right)
\]
where $\bar x^*=\sum_{i\in {\cal I}_{_N}} x^*_i$. 
\end{Pro}

\noindent{\bf Remark.}  In fact the following inequalities are satisfied by any equilibrium $x^*$:
\[
1-\frac{1}{\bar x^*} \varphi_i'\left(\frac{x^*_i}{\bar x^*}\right)>0, \quad i\!\in {\cal I}_{_N}.
\]
This follows from the convexity of the function $\xi \mapsto \xi -\varphi_i\left(\frac{\xi}{\bar x^*}\right)$ which is zero at $x_i^*$ with positive derivative.  As a consequence, the right hand side in  the above sufficient condition  is always positive which makes this criterion more realistic. 

\bigskip
\noindent {\bf Proof.}   Elementary computations show that  the differential $Dh(x)$ of $h$ at $x\!\in \R_+^N$ is given by
\[
\forall\, i,\, j\!\in {\cal I}_{_N},\qquad \frac{\partial h_i}{\partial x_j}(x)= \delta_{ij}\left(1-\frac{1}{\bar x} \varphi_i'\left(\frac{x_i}{\bar x}\right)\right) + \frac{x_i}{\bar x^2}\varphi_i'\left(\frac{x_i}{\bar x}\right).
\]
As a consequence all the diagonal terms of $Dh(x^*)$ are positive.   The above condition for all the eigenvalues of $Dh(x)$ to have positive real parts follows from a standard application of Gershgorin's Lemma to the transpose of $Dh(x)$.$\cqfd$

\subsection{A competitive system}

But once again, even if we could show that all equilibrium points are noisy traps, the convergence would not follow for free since this algorithm is associated to a so-called {\em competitive system}. A competitive differential system $\dot x= h(x)$ is a system in which the  field $h:\R^N\to \R^N$ is differentiable and satisfies
\[
\forall x\!\in \R^N ,\;\forall\, i,\, j\in {\cal I}_{_N},\; i\neq j,\; \quad \frac{\partial h_i }{\partial x_j}(x)>0 .
\]

As concerns Almgren and Harts's algorithm, the mean function $h$ is given by~(\ref{h2}), and under the standard differentiability assumption on the functions $\varphi_i$'s,
\[
\forall x\!\in \R^N ,\quad \frac{\partial h_i }{\partial x_j}(x) =\varphi'_i\left(\frac{x_i}{x_1+\cdots+x_{_N}}\right) \frac{x_i}{(x_1+\cdots+x_{_N})^2}>0.
\]

These systems are known to have possibly a non converging behaviour even in presence of a single (attracting) equilibrium. This is to be compared to  their {\em cooperative} counterparts (with negative non-diagonal partial derivatives) whose flow converge uniformly on compact sets toward the single equilibrium  in that case. This property can be transferred to the stochastic procedure by the mean if the so-called $ODE$ method which shows that the algorithm almost behaves like some trajectories of the Ordinary differential Equation  associated to its mean field $h$ (see $e.g.$~\cite{KUYI, BEMEPR,DUF} for an introduction). Cooperativeness and competitiveness are in fact some criterions which ensure some generalized monotonicity properties on the flow of the $ODE$ viewed as a function of its space variable. For some background on cooperative and competitive systems we refer to~\cite{ANHI, ANHI2} and the references therein.

\section{Numerical tests}\label{Sept}
\setcounter{equation}{0}
\setcounter{Assumption}{0}
\setcounter{Theorem}{0}
\setcounter{Proposition}{0}
\setcounter{Corollary}{0}
\setcounter{Lemma}{0}
\setcounter{Definition}{0}
\setcounter{Remark}{0}
\setcounter{figure}{0}



The aim of this section is to compare the behaviour
 of both algorithms on different data sets :  simulated i.i.d. data, simulated $\alpha$-mixing data and   (pseudo-)real data. 

Two natural situations of interest can be considered {\em a priori}:  {\em abundance} and {\em shortage}. By  ``abundance" we mean 
$\E V\leq \sum_{i=1}^N\E D_i$ (in average, the requested volume is   lower than the available one). The  "shortage" setting is the reverse situation where   $\E V> \sum_{i=1}^N\E D_i$. 

In fact, in the ``abundance" setting, both our procedures (optimization and reinforcement) tend to remain ``frozen" at their starting allocation value (usually uniform allocation) and they do not provide a significant improvement with respect to more naive approaches. By contrast the shortage setting is by far more commonly encountered on true markets and turns out to be much more challenging for our allocation procedures, so from now on we will focus on this situation.

Our first task is to define a reference strategy. To this end, we introduce an ``oracle strategy" devised by an insider who knows all the values $V^n$ and $D^n_i$ before making his/her optimal execution requests to the dark pools. It can be described as follows: assume for simplicity that the rebates are ordered $i.e.$ $\rho_1>\rho_2>\cdots>\rho_N$. Then, it is clear that 
%
%
the ``oracle" startegy yields the following cost reduction (CR) of the execution at time $n\ge 1$, 
$$
\hbox{CR}^{oracle} :=\left\{ \begin{array}{ll}
\displaystyle\sum_{i=1}^{i_0-1} \rho_iD^n_i+\rho_{i_0}\left(V^n-\sum_{i=1}^{i_0-1} D^n_i\right), & \mbox{if  }  \displaystyle\sum_{i=1}^{i_0-1} D^n_i\leq V^n<\sum_{i=1}^{i_0} D^n_i \\
\displaystyle\sum_{i=1}^N\rho_iD^n_i, & \mbox{if  }  \displaystyle\sum_{i=1}^N D^n_i< V^n.
\end{array}\right.
$$
Now, we introduce indexes to measure the performances of our recursive allocation procedures.
\begin{itemize}
	\item {\bf Relative cost reduction (w.r.t. the regular market):} they are defined as the  ratios between the cost reduction of the execution using dark pools and the   cost resulting from an   execution on the regular market for the three algorithms, $i.e.$, for every $n\ge 1$,
	\begin{itemize}
		\item[$\circ$] \mbox{\em Oracle:  }    \qquad $\displaystyle\frac{CR^{oracle}}{V^n}$ 		 		\item[$\circ$] \mbox {\em Recursive ``on-line" algorithms: }   \qquad $\displaystyle\frac{CR^{algo}}{V^n}= \frac{\sum_{i=1}^N\rho_i\min\left(r_i^nV^n,D_i^n\right)}{V^n}$ \\(with {\em algo} = {\em opti, reinf}). 
	\end{itemize}
			\item {\bf Performances (w.r.t. the oracle):}  the ratios between the relative cost reductions of our allocation  algorithms and that of the oracle, $i.e.$ for every $n\geq 1$
\[
\frac{CR^{opti}}{CR^{oracle}}\qquad \mbox{ and }\qquad \frac{CR^{reinf}}{CR^{oracle}}
\]
\end{itemize}

\noindent which seems a more realistic measure of the performance of our allocation procedures since the oracle strategy cannot be beaten.

Since these relative cost reductions are strongly fluctuating (with variables $V^n$ and $D^n_i$ in fact),  we will plot {\em the moving average} of these ratios (on the running period of interest)  and express them in pourcentage.

Moreover, when we simulate the data, we have chosen $10^4$ simulations because it corresponds approximatively to the number of pseudo-real data observed within a day. 

The choice of the gain parameter is the following (in the different settings considered below)
$$\gamma_n=\frac{c}{n}, \quad n\geq1$$
where $c$ equals to some units.

\subsection{The $(IID)$ setting}

We consider here simulated data in the i.i.d. setting, where the quantity $V$ and $D_i$, $i\in {\cal I}_N$, are log-normal variables and $N=3$. The variables $V$ and $D_i$, $i\in {\cal I}_N$, satisfy the assumptions of the CLT and we have the rate of convergence at least  of the optimization algorithm. 

%
The shortage setting is specified as follows: 
$$
\E V=\frac{3}{2}\sum_{i=1}^N\E D_i 
$$
with
$$
\E D_i=i, \ 1\leq i\leq N, \quad  \mbox{Var}(V)=1, \mbox{Var}(D_i)=1, 1\leq i\leq N \quad \mbox{and} \ \rho=\begin{pmatrix} 0.01 \cr 0.03 \cr 0.05 \end{pmatrix}.
$$
The running means of the performances are computed from the very beginning for the first 100 data, and by a moving average on a window of 100 data.

The initial value for both algorithms is set at  $r_i^0=\frac{1}{N}$, $1\leq i\leq N$. 
\begin{figure}[!h]
\centering
\vskip -0.25 cm
\epsfig{file=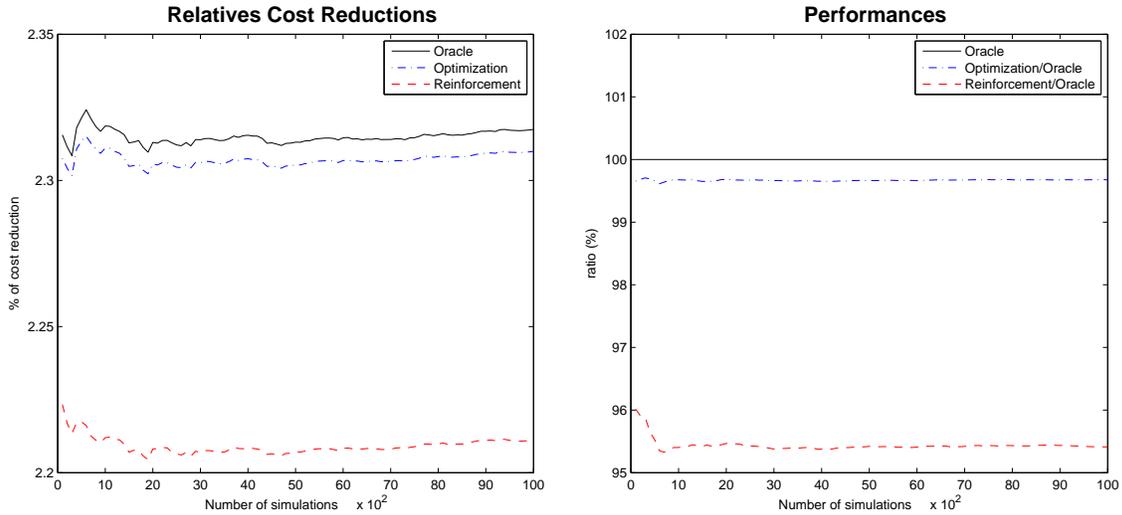,width=15cm}
\caption{\underline{\textit{Shortage setting}} Case $N=3$, $m_V=\frac{3}{2}\sum_{i=1}^Nm_{D_i}$, $m_{D_i}=i$, $\sigma_V=1$, $\sigma_{D_i}=1$, $1\leq i\leq N$.}
\label{FigIID}
\end{figure}

As expected, the optimization procedure outperforms  the reinforcement one and both procedures quickly converge (see~Figure~\ref{FigIID}) with respect to the data set size. Note that  the allocation coefficients (not reproduced here) generated by the two algorithms are significantly  different.  A more interesting feature is that the performances of the optimization procedure  almost replicate those of the  ``oracle".  Further simulations suggest that  the optimization algorithm   also  seems more robust when the variances of the random variables fluctuate.

\subsection{The $(ERG)$ setting}

We consider here simulated data in the ergodic setting, where the quantity $V$ and $D_i$, $i\in {\cal I}_N$, are exponentials of an Ornstein-Uhlenbeck process, $i.e.$
$$X^{n+1}=m+AX^n+B\Xi^{n+1},$$
where $\left\|A\right\|<1$, $BB^*\in GL(d,\R)$ and
$$m=\begin{pmatrix}m_1 \cr \vdots \cr m_{N+1}\end{pmatrix}\in{\R}^{N+1},   \quad \Xi^{n+1}=\begin{pmatrix}\Xi_1^{n+1} \cr \vdots \cr \Xi_{N+1}^{n+1}\end{pmatrix}\sim\mathcal{N}\left(0,I_{N+1}\right) \ \mbox{i.i.d.}, \quad e^{X^n}=\begin{pmatrix}V^{n} \cr D_1^{n} \cr \vdots \cr D_N^{n}\end{pmatrix}.$$

We are still interested in the shortage situation. The initial value of the algorithms is $r_i^0=\frac{1}{N}$, $1\leq i\leq N$ and we set  
$$\rho=\begin{pmatrix} 0.01 \cr 0.03 \cr 0.05 \end{pmatrix}, \quad
m=\begin{pmatrix} 1 \cr \vdots \cr 1 \end{pmatrix}, \quad A=\left(\begin{array}{cccc}
						0.7 & 0.01 & 0.01 & 0.01 \\
						0.01 & 0.3 & 0.01 & 0.01 \\
						0.01 & 0.01 & 0.2 & 0.01 \\
						0.01 & 0.01 & 0.01 & 0.1 \\
						\end{array}\right), \quad
B=\left(\begin{array}{cccc}
							0.02 & 0 & 0 & 0 \\
							0.01 & 0.9 & 0 & 0 \\
							0.01 & 0.01 & 0.6 & 0 \\
							0.01 & 0.01 & 0.01 & 0.3 \\
							\end{array}\right). \\$$
The running means of the performances are computed from the very beginning for the first 100 data, and by a moving average on a window of 100 data.

\begin{figure}[!h]
\centering
\epsfig{file=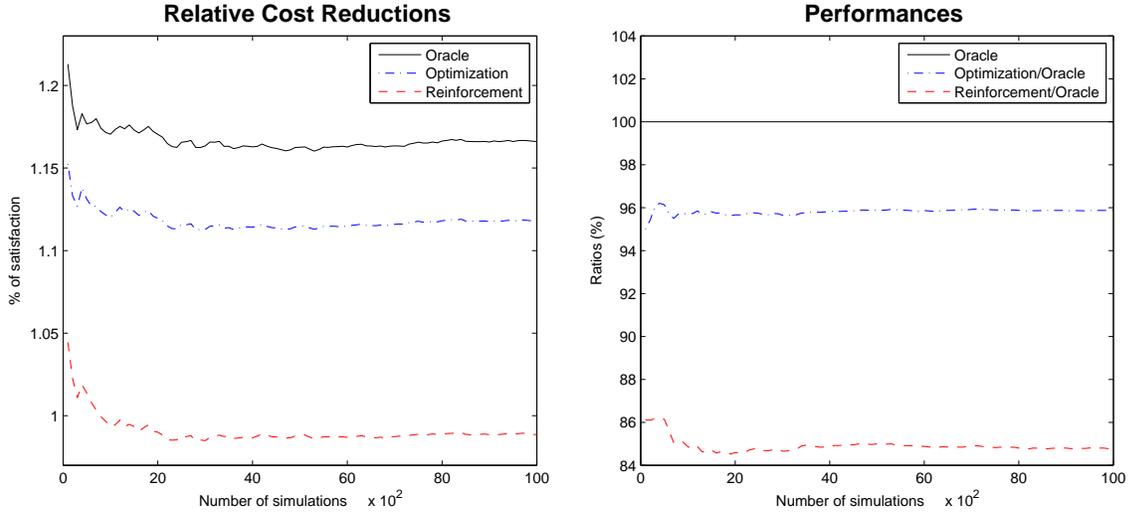,width=15cm}
\caption{\underline{\textit{Shortage setting}}: Case $N=3$, $m_V\geq \sum_{i=1}^N m_{D_i}$, $\sigma_V=1.21$, $\sigma_{D}=(8.21,\,3.05,\,1.07)^{'}$.}
\label{FigERG}
\end{figure}


We observe  in this ergodic setting a very similar behaviour to the i.i.d. one, with maybe a more significant advantage for the optimization approach (see~Figure~\ref{FigERG} right): the difference between the performances of both algorithms reaches 11\% in favour of the optimization algorithm.

\subsection{The pseudo-real data setting}

Firstly we explain how the data have been created. We have considered for $V$ the traded volumes of a very liquid security --~namely the asset BNP~--  during an $11$ day period. Then we selected the $N$ most correlated assets (in terms of traded volumes) with the original asset. These assets are denoted $S_i$, $i=1,\ldots,N$ and we considered their traded volumes during the same 11 day period.
Finally,  the available volumes of each dark pool $i$ have been modelled  as follows using   the mixing function
$$
\forall 1\leq i\leq N, \quad D_i:=\beta_i\left((1-\alpha_i)V+\alpha_i S_i \frac{\E V}{\E S_i}\right)
$$
where $\alpha_i, \  i=1,\ldots,N$ are the mixing coefficients, $\beta_i, \  i=1,\ldots,N$ some scaling parameters and $\E V$ and $\E S_i$ stand for the empirical mean of the data sets of $V$ and $S_i$.

The shortage situation corresponds to   $\sum_{i=1}^N \beta_i <1$ since it implies  $\E\left[\sum_{i=1}^N D_i\right] <\E V$. 
	

The simulations presented here have been made with  four dark pools  ($N=4$). Since the data used here covers $11$ days and it is clear that unlike the simulated data, these pseudo-real data are not stationary: in particular they are subject to daily changes of trend and volatility (at least).  To highlight this resulting changes in the response of the algorithms, we have specified the days by drawing vertical doted lines. The dark pool pseudo-data parameters are set to  
$$ \beta=\begin{pmatrix} 0.1 \cr 0.2 \cr 0.3 \cr 0.2 \end{pmatrix} \quad \mbox{and} \quad \alpha=\begin{pmatrix}  0.4 \cr 0.6 \cr 0.8 \cr 0.2 \end{pmatrix}
$$
and the dark pool trading (rebate) parameters are set to
$$
\rho=\begin{pmatrix}0.01 \cr 0.02 \cr 0.04 \cr 0.06\end{pmatrix}.
$$

The mean and variance characteristics of the data sets of  $(V^n)_{n\geq1}$ and  $(D_i^n)_{n\geq1}$, $i=1,\ldots,4$  are the following:
\begin{center}
\begin{tabular}{|c|c|c|c|c|c|}
\hline
  & $V$ & $D_1$ & $D_2$ & $D_3$ & $D_4$ \\
\hline
Mean &  955.42 & 95.54 & 191.08 & 286.63 & 191.08 \\
\hline
Variance &  $2.01\times10^6$ & $9.05\times10^3$ & $4.29\times10^4$ & $4.73\times10^5$ &  $5.95\times10^4$\\
\hline
\end{tabular}
\end{center}

Firstly, we benchmarked both algorithms on the whole data set ($11$ days) as though it were stationary without any resetting (step, starting allocation, etc.). In particular, the running means of the performances are computed from the very beginning for the first 1500 data, and by a moving average on a window of 1500 data. As a second step, we proceed on a daily basis by resetting the parameters of both algorithms (the initial profit for the reinforcement algorithm ($i.e.$ $I_i=0$, $1\leq i\leq N$) and the step parameter $\gamma_n$ of the optimization procedure) at the beginning of every day. The performances of both algorithms are computed on each day.

\paragraph{$\rhd$ Long-term optimization}  We observe that, except for the first and the fourth days where they behave similarly,  the optimization algorithm is more performing than the reinforcement one. Its performance is approximately 30\,\% higher on average (see Figure~3).  

\clearpage
\begin{figure}[!h]
\centering
\epsfig{file=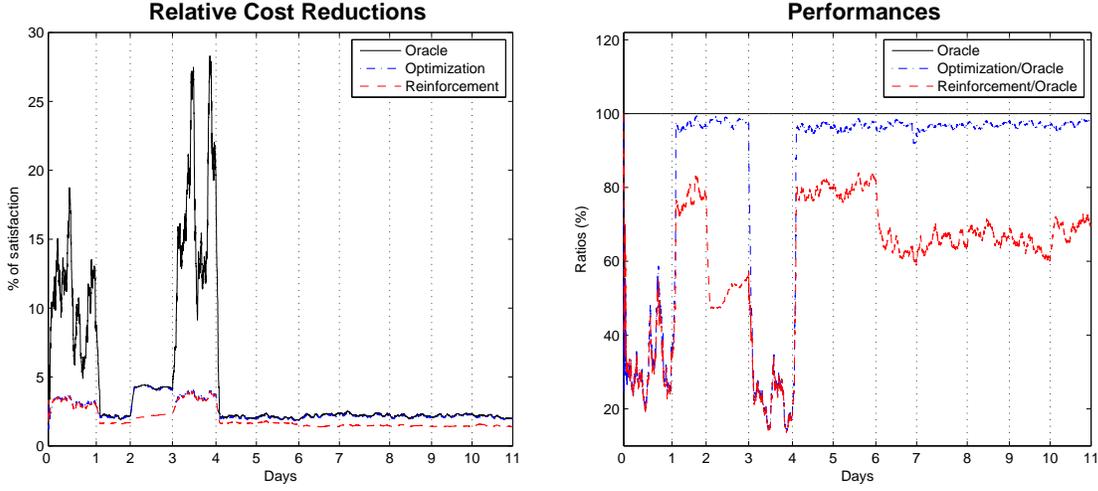,width=15cm}
\caption{\underline{\textit{Long term optimization}}: Case $N=4$, $\sum_{i=1}^N\beta_i<1$, $0<\alpha_i\leq0.2$ and $r^0_i=1/N$, $1\leq i\leq N$.}
\label{FigRealData}
\end{figure}
This test  confirms that the statistical features of the data are strongly varying from one day to another (see~Figure~\ref{FigRealData}), so there is no hope that our procedures converge in standard sense on a long term period. Consequently, it is necessary to switch to a short term monitoring by resetting the parameters of the algorithms on a daily basis as detailed below.


\paragraph{$\rhd$ Daily resetting of the procedure} We consider now that we reset each day all the parameters of the algorithm, namely we reset the step $\gamma_n$ at the beginning of each day and the satisfaction parameters and  we keep the allocation coefficients of the precedent day. We obtains the following results

\begin{figure}[!h]
\centering
\epsfig{file=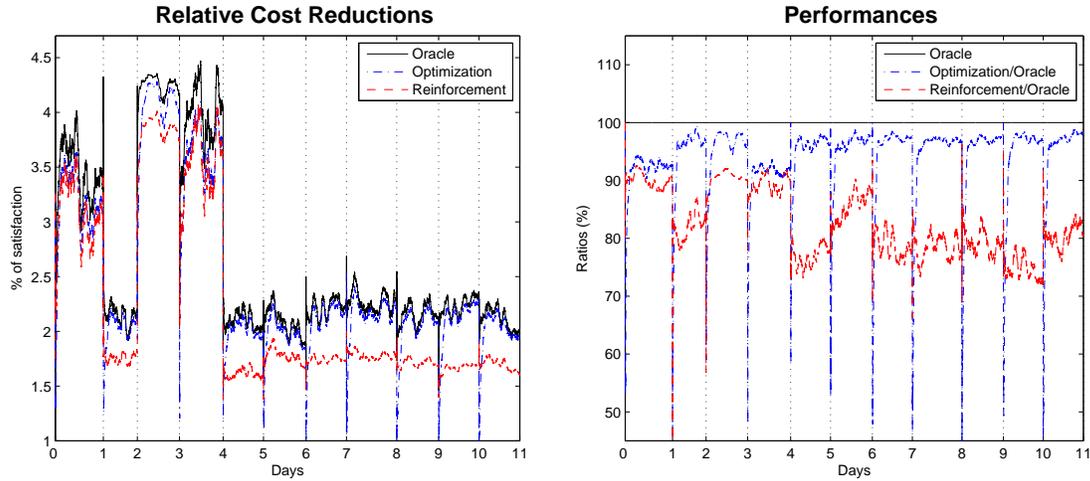,width=14.7cm}
\caption{\underline{\textit{Daily resetting of the algorithms parameters}}: Case $N=4$, $\sum_{i=1}^N\beta_i<1$, $0<\alpha_i\leq0.2$ and $r^0_i=1/N$ $1\leq i\leq N$.}
\label{Reset}
\end{figure}
\clearpage
We observe (see~Figure~\ref{Reset}) that the optimization algoritm still significantly outperforms the reinforcement one, reaching more $95\, \%$ of the performance of the oracle. Furthermore, although not represented here, the allocation coefficients  look  more stable.

\section{Provisional remarks}\label{Huit}
\subsection{Toward more general mean execution functions}
One natural idea is to take into account that the rebate may depend on the quantity $rV$ sent to be executed by the dark pool. The mean execution function of the dark pool can be modeled  by
\begin{equation}\label{varphi2}
\forall\, r\!\in [0,1],\qquad \varphi(r) = \E (\rho(rV)\min(rV,D))
\end{equation}
where the rebate function $\rho$ is a non-negative, bounded, non-decreasing right differentiable function.

For the sake of simplicity, we assume that $(V,D)$ satisfies~(\ref{noatom}).  The right derivative of $\varphi$ reads
\begin{equation}\label{Derivphi2}
\varphi_r'(r)=\E\left(\rho_r'(rV)V\min\left(rV,D\right)\right)+\E\left(\rho(rV)V\mbox{\bf 1}_{\left\{rV < D\right\}}\right),
\end{equation}
with in particular $\varphi'(0)= \rho(0)  \,\E ( V\mbox{\bf 1}_{\{D>0\}})>0$ as above. The main gap is to specify the function $\rho$ so that $\varphi$ remains concave which is the key assumption to apply the convergence theorem.  Unfortunately the choice for $\rho$ turns out to strongly depend on the (unknown) distribution of the random variable $D$. Let us briefly consider the case where $V$ and $D$ are independent  and $D$ has an exponential distribution ${\cal E}(\lambda)$.

First note that the function $g$ defined by $g(u):=\E (u\wedge D)  ,\, u\ge 0$ is given by 
\[
\forall\, u\ge 0, \quad g(u)= \frac{1-e^{-u\lambda}}{\lambda}
\]
so that, owing to the independence of $V$ and $D$,  
\[
\forall\, r\ge 0, \quad\varphi(r) = \E (\rho(rV) g(rV)).
\]
At this stage, $\varphi$ will be concave as soon as the function $\rho\, g$ is so. Among all possible choices, elementary computations show that a class of  possible choices is to consider $\rho= g^{\theta}$ with $\theta\!\in (0,\lambda]$. Of course this may appear as not very realistic since the rebate function is a structural feature of the different dark pools.

However several numerical  experiments not reproduced here testify that both  algorithms are robust to a realistic choice for the function $\rho$ $e.g.$ a non-decreasing and stepwise constant.

\medskip
Another natural extension is to model the fact that the dark pool may  take into account the volume $rV$ to decide which quantity will really executed rather than simply the {\em a priori} deliverable quantity $D$. One reason for such a behaviour is that  the dark pool may wish to preserve the possibility of future transactions with other clients. 

One way to model this phenomenon is to introduce a {\em delivery function} $\psi:\R_+^2\to \R_+$,  non-decreasing and concave   w.r.t. its first variable and satisfying $0\le \psi(x,y)\le y$, so that the new mean execution function is as follows: 
\begin{equation}\label{varphi3}
\varphi(r)= \rho\,\E\left(\min(rV,\psi(rV,D))\right).
\end{equation}
 It is clear that the function $\varphi$ is  concave (as the minimum of two concave functions) and bounded.
In this case, the first (right) derivative of $\varphi$ reads
\begin{equation}\label{Derivphi3}
\varphi_r'(r)=\rho \,\E\left(V\left(\mbox{\bf 1}_{\left\{rV <\psi(rV,D)\right\}}+ \psi_x'(rV,D)\mbox{\bf 1}_{\left\{rV\ge \psi(rV,D)\right\}}\right)\right)
\end{equation}
where $\psi'_x$ denotes the right derivative with respect to $x$. In particular $\varphi'_r(0)=  \rho   \,\E ( V\mbox{\bf 1}_{\{D>0\}})>0$.

\medskip 
As concerns the implementations resulting from  these new execution functions, the adaptation is straightforward. Note for the optimization procedure under constraints  that, firstly, the ``edge" functions  $R_i$ functions are not impacted by  the type of the execution function. On the other hand the definition of the  functions $H_i$ or the updating of the variables $I^n$ for the reinforcement procedure should be adapted in 
accordance with the representations~(\ref{Derivphi2}) and~(\ref{Derivphi3}) of 
the new mean  execution function $\varphi$.

\bigskip 
\noindent{\sc Example:} We consider for modelling the quantity delivered by the dark pool $i$ a function where we can define a minimal quantity required to begin to consum $D_i$, namely 
$$\psi_i(rV,D_i)=D_i \mbox{\bf 1}_{\{ rV> s_i D_i\}}$$
where $s_i$ is a parameter of the dark pool $i$ assumed to be deterministic. 

\bigskip


%



\noindent{\bf Pseudo-real data setting}
\begin{figure}[!h]
\centering
\epsfig{file=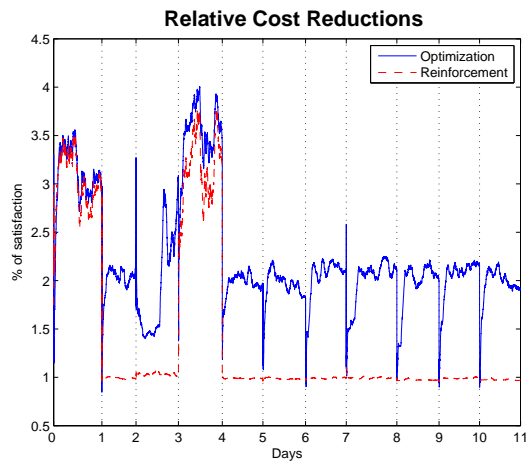,width=8cm}
\vskip
-0.5 cm
\caption{\underline{\textit{Shortage setting}}: Case $N=4$, $\sum_{i=1}^N\beta_i<1$, $0<\alpha_i\leq0.2$ and $r^0_i=1/N$, $1\leq i\leq N$, $s=\left(0.3,0.2,0.2,0.3\right)^t$.}
\end{figure}

\subsection{Optimization $vs$ reinforcement ?}
For practical implementation what conclusions can be drawn from our investigations on both procedures. Both reach quickly a stabilization/convergence  phase close to optimality. The reinforcement algorithm leaves the simplex structurally stable which means the proposed dispatching at each time step is  realistic whereas the  stochastic Lagrangian algorithm in its present form  sometimes needs to be corrected from time to time. This can be corrected by adding a  projection on the simplex at each step. We did not consider this variant from a theoretical point of view to  keep our convergence proofs more elementary. 

In a high volatility context, the stochastic Lagrangian algorithm clearly  prevails  with performances that turn out to be     significantly better.  This optimization procedure also relies on established convergence results in a rather general framework (stationary $\alpha$-mixing input data). 
However, given the computational cost of these procedures which is close to zero, a possible strategy is  to implement them in parallel to get a synergistic effect. In particular, one may the reinforcement algorithm --~which step parameter  is structurally fixed equal to~$\frac 1n$~-- can be used to help tuning the constant $c$ in the gain parameter $\gamma_n = \frac c n$ of the stochastic Lagrangian. Doing so one may start with a small constant $c$, preventing a variance explosion of the procedure. Then based one may increase slowly this constant until the Lagragian outperforms the reinforcement procedure.   



\appendix

\begin{center}
\Large{\textbf{Appendix}}
\end{center}
\section{\large{Robbins-Zygmund with averaging innovation}}
\setcounter{equation}{0}
\setcounter{Assumption}{0}
\setcounter{Theorem}{0}
\setcounter{Proposition}{0}
\setcounter{Corollary}{0}
\setcounter{Lemma}{0}
\setcounter{Definition}{0}
\setcounter{Remark}{0}

\medskip
We consider an algorithm of the following form
\begin{equation}\label{AlgoSto}
\theta_{n+1}=\theta_n-\gamma_{n+1}(G(\theta_n, Y_{n})+\Delta M_{n+1}), \quad n\geq 0
\end{equation}
where $G$ is a Borel function from ${\R}^d\times{\R}^q$ to ${\R}^d$, $(Y_n)_{n\geq0}$ is a sequence of ${\R}^q$-valued 
 random vectors  adapted to a filtration $({\cal F}_n)_{n\ge 1}$, $\theta_0$ is an ${\cal F}_0$-measurable $\R^d$-valued random vector independent of $(Y_n)_{n\ge 1}$, all
defined on the same probability space $\left(\Omega,{ \cal F},\P\right)$,  $(\Delta M_n)_{n\ge 1}$ is a sequence of   ${\cal F}_n$-martingale increments and  $(\gamma_n)_{n\geq1}$ is a non-increasing sequence of positive real numbers going to $0$ as $n$ goes to infinity.  

\bigskip
We will say that  $(Y_n)_{n\geq0}$ is 
$\nu$-averaging (under ${\P}$) on a class of functions ${\cal V}_{0^+,p}\subset L^p(\nu)$ if, for every $p\!\in [1,+\infty)$, 
\begin{equation}\label{Ergodic}
\forall f\in {\cal V}_{0^+,p}, \quad\frac{1}{n}\sum_{k=0}^{n-1}f(Y_k)\longrightarrow\int_{{\R}^q}fd\nu\quad \P\mbox{-}a.s. \;\mbox{ and in }\; L^p(\P).
\end{equation}


\noindent Let $\beta\!\in (0,1)$, let $p\in[1,\infty)$. We denote by $\mathcal{V}_{\beta,p}$ the class of functions whose convergence rate in~(\ref{Ergodic}) $\P\mbox{-}a.s.$ and in $L^p(\P)$ is $O(n^{-\beta})$, namely 
\begin{equation}
	\mathcal{V}_{\beta,p}=\left\{f:{\R}^q\rightarrow{\R} \left|\right. \frac{1}{n}\sum_{k=1}^{n}f(Y_k)-\int fd\nu\stackrel{\P\mbox{-}a.s. \ \& \ L^p(\P)}{=}O(n^{-\beta})\right\}.
\label{ClasseBetaP}
\end{equation}

Now we are in a position to state the convergence theorem.

\medskip
\noindent {\bf Theorem~A.1} 
{\it (A Robbins-Zygmund like Lemma) Let  $G : {\R}^d\times {\R}^q \rightarrow  {\R}^d$ be a Borel function, let $(Y_n)_{n\geq0}$ be an ${\cal F}_n$-adapted $\nu$-averaging   sequence of $\R^q$-valued random vectors and let $(\Delta M_n)_{n\ge 1}$ be a sequence of   ${\cal F}_n$-martingale increments. Assume that there exists a continuously differentiable function $L: {\R}^d\rightarrow  {\R}_+$ satisfying
\begin{equation}\label{Lyapunov}
	\nabla L \ is \ Lipschitz \ continuous \ and \ \left|\nabla L \right|^2\leq C\left(1+L\right)
\end{equation}
such that the function $G$
satisfies the following \textnormal{local weak mean-reverting} assumption:  
\begin{equation}\label{lmr}
 \forall\, \theta\!\in \R^d,\; \forall\,y\!\in \R^q,\quad  \left\langle \nabla L(\theta)\left|\right. G(\theta,y)-G(\theta^*,y)\right\rangle \ge 0.
 \end{equation}

Suppose there exists $\beta\!\in (0,1)$, $p\!\in [1,\infty)$  such that
\begin{equation}\label{GH2}
	 G(\theta^*,\cdot)\in\mathcal{V}_{\beta,p}.
\end{equation}
Moreover, assume that $G$ satisfies the following linear growth assumption
\begin{equation}
	\forall \theta\in{\R}^d, \forall y\in{\R}^q, \quad \left|G(\theta,y)\right| \leq  \varphi(y)(1+L(\theta))^{\frac 12}\label{HH} 
\end{equation}
and
\begin{equation}
	\E(|\Delta M_{n+1}|^2\,|\, {\cal F}_n)\leq   \varphi^2(Y_n)(1+L(\theta_n))\label{MH} 
\end{equation}
where the function  $ \varphi$ satisfies $\sup_n \| \varphi(Y_n)\|_{2\vee\frac{p}{p-1}}<+\infty$ (convention $\frac{1}{0}=+\infty$).

Let $\gamma=(\gamma_n)_{n\geq1}$ be a non-negative, non-increasing sequence of gain parameters satisfying
\begin{equation}\label{gamma}
	\sum_{n\geq1}\gamma_n=+\infty, \quad n^{1-\beta}\gamma_n\longrightarrow 0, \quad \mbox{and} \quad \sum_{k\geq1}k^{1-\beta}\max\left(\gamma_k^2,\left|\Delta\gamma_{k+1}\right|\right)<+\infty.
\end{equation}
Assume that $\theta_0$ is ${\cal F}_0$-adapted and $L(\theta_0)<+\infty$, $\P$-$a.s$. Then, the recursive procedure defined by~(\ref{AlgoSto}) satisfies 
$$
L(\theta_n)\stackrel{a.s.} {\longrightarrow} L_\infty<+\infty\; \mbox{ and }\; \sum_{n\ge 0} \g_{n+1}\langle \nabla L(\theta_n)\,|\, G(\theta_n,Y_n)-G(\theta^*,Y_n)\rangle<+\infty \; a.s.
$$
}


\small
\begin{thebibliography}{0}
\bibitem
{OPTEXECAC00} {\sc Almgren, R.~F. and Chriss N.} (2000).
\newblock Optimal execution of portfolio transactions.
\newblock {\em Journal of Risk}, 3(2):5-39.
\bibitem
{almsor08}
{\sc Almgren R. and Harts B.} (2008).
\newblock A dynamic algorithm for smart order routing.
\newblock Technical report, StreamBase.
\bibitem{BEMEPR} {\sc Benveniste M., M\'etivier M., Priouret P.} (1987).  {\em Adaptive Algorithms and Stochastic Approximations}, Springer Verlag, Berlin, 365p.                         
\bibitem{BELE} {\sc Berenstein R.} (2008).  {\em Algorithmes stochastiques, microstructure et ex\'ecution d'ordres}, Master 2 internship report (Quantitative Research, Dir. C.-A. Lehalle, CA Cheuvreux), {\em Probabilit\'es \& Finance}, UPMC-\' Ecole Polytechnique.
\bibitem{DEDetal} {\sc Dedecker J., Doukhan P., Lang G., Le\'on J.R., Louhichi S., Prieur C.} (2007).  {\em Weak dependence} (with Examples and Apllications), Lecture notes in Statistcs, {\bf 190}, Springer, Berlin, 317p.
 \bibitem{DOU}  {\sc Doukhan P.} (1991).  {\em Mixing: Properties and Examples}, Lect. Notes Statist. {\bf 85},  Springer, New York, 142p. 
   \bibitem{DUF}  {\sc Duflo M.} (1997). {\em Random Iterative Models}. Series: Stochastic Modelling and Applied Probability, Vol. 34. XVI, 385 p.,  Springer, Berlin.
   \bibitem
   {FOU06} {\sc Foucault, T. and Menkveld, A.~J.} (2006).
\newblock Competition for order flow and smart order routing systems, {\em Journal of Finance}, {\bf 63}(1):119-158.
\bibitem{GER} {\sc Gantmacher F. R. }(1959). {\em The theory of matrices}, {\bf 1-2}, Chelsea, New York, 374p.
  \bibitem{GAKO} {\sc G\'al I.S. and Koksma J.F.} (1950). Sur l'ordre de grandeur des fonctions sommables, {\em Indigtiones Math.}, {\bf 12}:192-207.
\bibitem{ANHI}{\sc  Hirsch  M.,  Smith H.} (2004). {\em Monotone Dynamical systems}, monography, 136p.
\bibitem{ANHI2}{\sc  Hirsch  M.,  Smith H.} (2005). {\em Monotone Dynamical systems}, Handbook of differential equations: ordinary differential equations. Vol. II, Elsevier B. V., Amsterdam, 2005, 239-357.
 \bibitem{KUYI} {\sc Kushner H.J., Yin G.G.} (1997). Stochastic approximation and recursive algorithms and applications, New York, Springer, 496p; and $2^{nd}$ edition, 2003. 
\bibitem{LAPATA}{\sc Lamberton D., Pag\`es G., Tarr\`es P.}  (2004). When can the two-armed bandit algorithm be trusted? {\em The Annals of Applied Probability}, {\bf 14}(3):1424-1454.
\bibitem{LAPA1}{\sc Lamberton D., Pag\`es G.} (2008). How fast is the bandit? {\em Stoch. Anal. and Appl.}, {\bf 26}:603-623.
\bibitem{LAPA2}{\sc Lamberton D., Pag\`es G.} (2008) A penalized bandit algorithm, {\em EJP}, {\bf 13}:341-373.
\bibitem{LAR}   {\sc Laruelle S.}, PhD Thesis, in progress.
\bibitem{LARPAG}   {\sc Laruelle S., Pag\`es G.} (2009). Stochastic Approximation with Ergodic Innovation, in progress. 
 \bibitem
 {citeulike:5094012} {\sc  Lehalle C.-A.} (2009).
\newblock Rigorous strategic trading: Balanced portfolio and mean-reversion.
\newblock {\em The Journal of Trading}, {\bf 4}(3):40-46.
   \bibitem{LEPA}  {\sc Lemaire V., Pag\`es G.} (2008).  Unconstrained recursive importance sampling, pre-pub LPMA 1231, to appear in {\em Annals of Applied Probability}.
   \bibitem{PAG}  {\sc Pag\`es G.} (2001).  Sur quelques algorithmes r\'ecursifs pour les probabilit\'es num\'eriques, {\em ESAIM P\&S}, {\bf 5}:141-170.
\bibitem{TAR}{\sc Tarr\`es P.}  (2001). Traps of stochastic algorithms and vertex-reinforced random walks, th\`ese d'Universit\'e, ENS Cachan, 180p.
 \end{thebibliography}
 \end{document}